# Giant Spin-to-Charge Conversion by Tailoring Magnetically Proximitized Topological Dirac Semimetal

*Masayuki Ishida[1], Soichiro Fukuoka[1], Takahiro Chiba[2,3]†††, Yohei Kota[4], Masaaki Tanaka[1,5,6]†† and Le Duc Anh[1,5]†*

[1]Dept. of Electrical Engineering and Information Systems, The University of Tokyo, Tokyo, Japan

[2]Department of Information Science and Technology, Graduate School of Science and Engineering, Yamagata University, Japan

[3]Department of Applied Physics, Graduate School of Engineering, Tohoku University, Japan

[4]National Institute of Technology, Fukushima College, Japan

[5]Center for Spintronics Research Network, The University of Tokyo, Tokyo, Japan

[6]Institute for Nano Quantum Information Electronics, The University of Tokyo, Tokyo, Japan

**ABSTRACT**.

The role of magnetic proximity on topological surface states (TSSs) and spin-to-charge conversion efficiency is studied in Fe/topological Dirac semimetal α-Sn heterostructures. As α-Sn thickness increases (9–38 nm), the damping constant of Fe peaks at 25 nm and then decreases. Theoretical modeling links this to long-range magnetic proximity effects (MPE) that suppress TSSs at the Fe/α-Sn interface while opening an exchange gap at the opposite interface. At 25 nm, we achieve an inverse Edelstein length of -3.1±1.0 nm, one of the highest values at room temperature, highlighting MPE's importance in spintronics.

†Contact author: anh@cryst.t.u-tokyo.ac.jp

††Contact author: masaaki@ee.t.u-tokyo.ac.jp

†††Contact author: t.chiba@yz.yamagata-u.ac.jp

Control and detection of magnetic order using spin current is promising for SOT-MRAM and power-efficient logic devices (e.g., MESO) [1–4]. For this purpose, topological materials are attractive since their topological surface states (TSS) can enable efficient spin–charge interconversion up to room temperature [5–9]. However, when interfaced with a ferromagnet (FM), which is the common structure in many spintronic devices, the FM's magnetization breaks time-reversal symmetry (TRS) and can modify or even destroy these TSSs (e.g., opening an exchange gap) [5,10,11]. Understanding how magnetic proximity alters the TSS—and thus the conversion efficiency—is essential for both fundamentals and device design.

To preserve the spin-momentum-locked TSSs, a commonly used technique is to insert a conductive spacer layer between the FM and the topological layers. For example, it has been shown that spin-charge interconversion is absent at the interface of Fe and topological insulator (TI) α-Sn thin film (5 nm). Meanwhile, an Fe/Ag (2 nm)/α-Sn (5 nm) heterostructure exhibited a very high spin-to-charge conversion efficiency, which was quantified as an inverse Edelstein length ($\lambda_{IEE}$) of 2.1 nm [5]. Similar highly efficient spin-to-charge conversion ($\lambda_{IEE}$ =2.0 [12] ~ 2.7 nm [7]) was also reported in other systems utilizing a spacer layer. However, inserting a spacer holds some inherent problems such as current shunting and increased possibility of spin scattering due to the additional layer, which hinders the efficiency of magnetization manipulation using charge-to-spin conversion.

In this letter, we aim to elucidate the evolution of TSS and the spin-to-charge conversion efficiency of topological materials under the influence of the MPE, using spacer-layer-less bilayers composed of a FM metal (Fe) and the topological Dirac semimetal (TDS) α-Sn [13–16] (see Fig. 1(a)). TDSs represent a rare class of topological materials characterized by linear band dispersions (Dirac cones) arising from band inversion and high crystalline symmetry in their bulk electronic structure [17,18]. These Dirac cones are composed of spin-non-degenerate linear bands, often referred to as Weyl cones, and exhibit strong coupling between spin, orbital, and momentum degrees of freedom [19]. Due to the presence of band inversion, TDSs also host TSSs on both sides of their thin films, similar to the case of TIs. These features make TDSs promising candidate materials for highly efficient spin-charge interconversion. However, unlike TIs, studies on spin transport and spin-charge interconversion in TDSs remain limited and underdeveloped [20,21]. In α-Sn, the TDS of choice in this study, promising preliminary results were reported to exhibit highly efficient spin-charge interconversion [5,6] and magnetization reversal [22], although the underlying mechanisms associated with its topological band structure have yet to be fully explored. The TSSs of α-Sn films were studied by photoemission and transport measurements [5,13,14,16,23–25], where the spin-polarization in TSS was also reported [16].

For our experiment, Fe (4 nm) / α-Sn ($t_{\mathrm{Sn}}$ = 0, 9, 25, 32.5, 38 nm) heterostructures were grown on an InSb substrate by molecular beam epitaxy (MBE), which will be referred to as S0, S9, S25, S32.5, and S38 for convenience. The details of sample growth and characterization can be found in Fig. 1(b-d) and Supplemental Material [26]. Frequency-dependent ferromagnetic resonance (FMR) spectroscopy was carried out on the samples at room temperature using a coplanar waveguide and Keysight vector network analyzer (VNA). A pronounced increase in Gilbert damping constant α is observed as the α-Sn layer thickness increases from $t_{\mathrm{Sn}}$ = 0 nm (no α-Sn layer, α ~ 0.003) to $t_{\mathrm{Sn}}$ = 25 nm (α ~ 0.0226), followed by a decrease as $t_{\mathrm{Sn}}$ exceeds 25 nm, as shown in the inset of Fig. 2(c). The increase in $\alpha$ with the insertion of the α-Sn layer indicates that the α-Sn layer or its surface acts as the spin sink.

Spin-pumping measurements were then carried out on the same series of samples at room temperature in a cavity-based ESR machine at ~ 9.1GHz in the setup shown in Fig. 2(a). For S25, as shown in Fig. 2(b), a clear FMR signal was observed at each microwave power, which did not show any saturation up to 100 mW [26]. Simultaneously, a dc voltage peak was observed between the two edges of the sample at the FMR field at each microwave power, as shown in Fig. 2(d). In order to exclude the contribution from the conventional Seebeck effect, which is independent of the direction of the applied magnetic field, we defined the effective voltage induced by spin-to-charge conversion in α-Sn as $V_{\mathrm{eff}} = (V_{H+} - V_{H-})/2$, where $V_{H+}$ and $V_{H-}$ are the voltages measured in opposite directions of the applied magnetic field $\boldsymbol{H}$, along $[\bar{1}10]$ and $[1\bar{1}0]$. This setup was determined as the one that shows a positive sign in Pt / Permalloy heterostructure, which is known to possess a positive inverse spin Hall angle. The $V_{\mathrm{eff}}$ linearly scaled with the microwave power, as shown in the inset of Fig. 2(d), which indicates successful spin-pumping-induced spin-to-charge conversion in S25. Surprisingly though, the sign of the effective voltage peak in S25 was negative. The origin of this negative sign is discussed later.

The voltage signal is more clearly observed as the magnitude of damping enhancement increases, indicating that the damping enhancement is positively correlated to the injected spin current. For example, no clear linear power dependency of the induced voltage peak was observed in S0 and S9, where the damping enhancement was among the lowest of the samples [26]. The fact that there are no signs of spin-to-charge conversion in S0 and S9 means that either the spin-to-charge conversion was not carried out or the generated charge current was too small compared to thermal noise due to the small injected spin current. Meanwhile, the induced voltage peak showed an increasing trend as a function of the RF power in S32.5 and S38, where the damping enhancement was larger than that of S0, and the clear linear dependence was observed in S25, where the damping enhancement was the largest. However, since the voltage signal of neither S32.5 nor S38 scaled linearly with increasing power, the voltage signal is concluded to have originated from parasitic effects such as spin-Seebeck and spin rectification effect (see Supplemental Material [26]).

There are four possible spin-to-charge conversion mechanisms in S25: Inverse Rashba-Edelstein effect at (a) the Fe/α-Sn interface (top TSS) and (b) the α-Sn/InSb interface (bottom TSS), and inverse spin Hall effect in (c) the α-Sn layer and (d) the InSb layer. The InSb layer contribution can be excluded due to the experimental fact that no spin-to-charge conversion was observed in S0. The Fe/α-Sn interface contribution can also be excluded because of the absence of spin-to-charge conversion in S9, S32.5, and S38, along with the Fe (5 nm) / α-Sn (5 nm) sample in Ref. [5]. Moreover, the absence of TSS at the Fe/α-Sn interface is supported by the previous work of Rojas-Sánchez *et al.* [5], where *in-situ* ARPES results of an Fe (5 nm)/α-Sn (5 nm)/InSb heterostructure—structurally similar to our samples—clearly showed the disappearance of the α-Sn surface Dirac cone when an ultrathin 0.18 nm-thick Fe was deposited. The α-Sn layer contribution is also unlikely, since the α-Sn-thickness-dependent evolution of the Gilbert damping constant did not saturate but rather showed a local maximum as the Sn thickness increased, suggesting that the dominant spin sink is not in the bulk state of α-Sn.

Only the scenario in which the contribution of the α-Sn/InSb interface is dominant is consistent with all the experimental observations. First, both the decrease in damping enhancement for α-Sn thicknesses above 25 nm and the observation of a detectable spin-pumping signal only in S25 can be accounted for by the theoretical model proposed later in this manuscript. Furthermore, the inverse Edelstein length of S25 has a negative sign, whereas a positive value was previously reported for an Fe/Ag-spacer/α-Sn structure, in which spin-to-charge conversion was attributed to the TSS at the upper α-Sn interface protected by the Ag spacer [5]. This sign reversal can be naturally explained only by invoking the TSS at the α-Sn/InSb interface: Since the spin-momentum-locking helicities of the top and bottom TSSs of topological Dirac semimetal α-Sn are expected to be opposite, as inferred in analogy with BiSbTe [27], spin-to-charge conversion mediated by the bottom surface state in S25 should have the opposite sign to that previously observed for the top surface state.

To investigate the evolution of the topological surface states (TSSs) of Fe/α-Sn heterostructures, we first carried out first-principles band structure calculations. Fig. 3 displays the calculated band structures for various α-Sn thicknesses interfaced with a single atomic layer of Fe whose magnetization is pointed perpendicular to the film plane. Two key observations emerge from these calculations. First, Dirac-cone-like TSS at the top Fe/α-Sn interface is largely destroyed due to the MPE and strong orbital hybridization with the Fe layer, leaving only surface states at higher energy or *k*-momentum as shown by the green dispersion curves in Fig. 3(b-d). This finding is consistent with the previous study [5] and underscores the significant influence of the MPE from the Fe layer on the TSS. Second, a gapped TSS is observed at the bottom α-Sn/InSb interface (modeled as the α-Sn/vacuum interface in Fig. 3(a)) as shown by the purple dispersion surrounded by dashed curves in Fig. 3(b-d). The gap energy of the TSS decreases as the α-Sn layer thickness increases and becomes zero above ~6.5 nm (40ML). Two possible mechanisms may account for the energy gap observed in the bottom TSS. The first is hybridization between the states on the opposing surfaces of α-Sn, as reported in previous studies [28–30]. However, in our system, the destruction of the top-surface TSS at the Fe/α-Sn interface should suppress this top–bottom surface state hybridization. This suggests that the second mechanism, the long-range MPE [10,11] from Fe, should be the main driver of the gap opening in the bottom TSS at the α-Sn/InSb interface.

From a microscopic perspective, there are two possible mechanisms through which the MPE from Fe can influence the bottom topological surface state (TSS). First, the electronic states of Fe, particularly its spin-polarized *d* orbitals, can hybridize with those of the α-Sn layer. Owing to the bulk–boundary correspondence linking the TSS to the bulk band topology, such spin-dependent hybridization can break time-reversal symmetry and induce an exchange gap in the TSS. Second, the long phase-coherence length of electrons in our α-Sn films may extend the effective range of the MPE across the layer. In our highest-quality sample, the phase-coherence length of the heavy-hole carriers reaches approximately 40 nm [31], which is comparable to or greater than the α-Sn thickness considered here. This long-range coherence may therefore enable the magnetic influence of the Fe layer to reach the bottom α-Sn/InSb interface and open a gap in the bottom TSS. Taken together, these considerations support the MPE as the most plausible origin of the gap opening in the bottom TSS.

Then, we developed an analytical model similar to that reported in Ref. [32] to explain the observed α-Sn layer thickness dependence of the damping constant enhancement. In the present model, spin pumping from the Fe layer generates spin accumulation $\boldsymbol{\mu}_{\mathrm{S}}(z)$ and spin current $\boldsymbol{J}_{\mathrm{S}}(z)$ which propagate along the growth direction ($z$ direction) in the bulk state of the α-Sn layer, as illustrated in Fig. 4(a). It should be noted that the direction of vectors $\boldsymbol{\mu}_{\mathrm{S}}$ and $\boldsymbol{J}_{\mathrm{S}}$ represent the direction of spin along the $y$ axis and not the flow along the $z$ axis. Note that this model is applicable because the TSS is absent at the top Fe/α-Sn interface, as revealed by the first-principles calculations and previous study [5]. The α-Sn layer is treated as a spin channel with an electrical conductivity $\sigma$ and a spin diffusion length $\lambda_{\mathrm{Sn}}$. At the bottom α-Sn/InSb interface ($z = t_{\mathrm{Sn}}$), the spin accumulation $\boldsymbol{\mu}_{\mathrm{R}}$ is determined by the inverse Rashba-Edelstein effect on the spin-momentum-locked TSS, which serves as a boundary condition for $\boldsymbol{\mu}_{\mathrm{S}}(z = t_{\mathrm{Sn}})$. Under the influence of MPE from the adjacent Fe layer, we assume that the TSS at the α-Sn/InSb interface acquires an exchange gap, resulting in a gapped two-dimensional (2D) Dirac-like dispersion described by $E_{\pm}(\boldsymbol{k}) = \pm\sqrt{(hv_{\mathrm{F}}\boldsymbol{k})^2 + \Delta^2}$, where $h$ is the Planck constant, $v_{\mathrm{F}}$ is the Fermi velocity of the ungapped TSS, and $\Delta$ is half of the exchange gap. The α-Sn thickness dependence of gap $\Delta$ was roughly modeled as $\Delta \propto 1/t_{\mathrm{Sn}}{}^2$ (see Ref. [29]).

Under this gapped condition, spin accumulation at the bottom α-Sn/InSb interface can be estimated as

$$\boldsymbol{\mu}_{\mathbf{R}} = -\boldsymbol{\mu}_{\mathbf{R}}^{\mathbf{0}} \frac{1 - \left(\frac{\Delta}{E_{\mathrm{F}}}\right)^2}{1 + 3\left(\frac{\Delta}{E_{\mathrm{F}}}\right)^2} \tag{1}$$

where $\boldsymbol{\mu}_{\mathrm{R}}^{0} = A\frac{\hbar}{2}\frac{2e\boldsymbol{E}_{\mathrm{y}}}{hv_{\mathrm{F}}}\frac{E_{\mathrm{F}}\tau_e}{\hbar}$, $A$ is the area of the α-Sn/InSb interface, $E_{\mathrm{F}}$ is the Fermi energy, $\boldsymbol{E}_{\mathbf{y}}$ is the generated electrical field, $h$ is the Planck constant, and $\tau_e$ is the momentum relaxation time of electrons at the TSS [26,33,34].

The damping enhancement $\delta\alpha$ is positively correlated to the effective spin current $\boldsymbol{J}_{\mathrm{eff}}$ (= $\boldsymbol{J}_{\mathrm{P}} - \boldsymbol{J}_{\mathrm{B}}$) injected into the α-Sn layer, where $\boldsymbol{J}_{\mathrm{P}}$ is the pumped spin current induced by the RF magnetic field and $\boldsymbol{J}_{\mathrm{B}}$ is the backflow spin current caused by the spin accumulation $\boldsymbol{\mu}_{\mathrm{s}}(z = 0)$ at the Fe/α-Sn interface. From the boundary conditions at the top Fe/α-Sn interface ($z = 0$) and the bottom α-Sn/InSb interface ($z = t_{\mathrm{Sn}}$), the spin accumulation $\boldsymbol{\mu}_{\mathrm{s}}(0)$ in the α-Sn layer can be described as

$$\boldsymbol{\mu}_{\mathrm{s}}(0) = \boldsymbol{J}_{\mathbf{eff}} \frac{e\lambda_{\mathrm{Sn}}}{\sigma} \tanh\left(\frac{t_{\mathrm{Sn}}}{\lambda_{\mathrm{Sn}}}\right) + \frac{\boldsymbol{\mu}_{\mathbf{R}}}{\cosh\left(\frac{t_{\mathrm{Sn}}}{\lambda_{\mathrm{Sn}}}\right)} \tag{2}$$

Then, as presented in detail in Supplemental Material [26], the relation between the damping enhancement $\delta\alpha$ and the α-Sn layer thickness $t_{\mathrm{Sn}}$ is expressed as

$$\delta\alpha = \frac{G_r \gamma \hbar^2}{2e^2 M_{\mathrm{s}} d_{\mathrm{M}}} \frac{1 + \frac{|\boldsymbol{\mu}_{\mathbf{R}}|}{\hbar} \operatorname{sech}\left(\frac{t_{\mathrm{Sn}}}{\lambda_{\mathrm{Sn}}}\right)}{1 + \frac{\lambda_{\mathrm{Sn}} G_r}{\sigma} \tanh\left(\frac{t_{\mathrm{Sn}}}{\lambda_{\mathrm{Sn}}}\right)} \tag{3}$$

where $\gamma$ is the gyromagnetic ratio of Fe, $\hbar$ is the Dirac constant, $d_{\mathrm{M}}$ and $M_{\mathrm{s}}$ are the thickness and saturation magnetization of the Fe layer, respectively.

From these equations, it is evident that the spin accumulation $\boldsymbol{\mu}_{\mathrm{R}}$ at the bottom α-Sn/InSb interface—governed by the gap opening in the TSS—directly influences the spin accumulation and, consequently, the backflow spin current at the top Fe/α-Sn interface. As a result, the exchange gap in the TSS at the bottom α-Sn/InSb interface affects the damping constant of the top Fe layer at the opposite interface. The red curve in Fig. 4(b) shows the fitting to the experimental data using Eq. (3) (see the fitting parameters in Table S-II in Supplemental Material [26]). Based on this analysis, the spin diffusion length ($\lambda_{\mathrm{Sn}}$) of α-Sn is estimated to be approximately 20 ± 5 nm, which is 3–4 times longer than that of typical 3D topological insulators such as $Bi_2Se_3$ [35] and BiSb [36]. It should be noted that the monotonic decrease of $\delta\alpha$ for $t_{\mathrm{Sn}} > 25$ nm is nearly identical in both the red and blue curves of Fig. 4(b), irrespective of the thickness dependence of the exchange gap Δ. As seen in Eq. (3), this decrease is governed by the hyperbolic functions, from which the spin diffusion length $\lambda_{\mathrm{Sn}}$ in bulk α-Sn is extracted. Consequently, essentially the same value of $\lambda_{\mathrm{Sn}} = 20 \pm 5$ nm is obtained regardless of whether our model assumes that Δ depends on the thickness or not.

By combining our first principles calculations and analytical model, the experimental observation in this study can be understood intuitively: When a spin-momentum-locked TSS is preserved at the bottom α-Sn/InSb interface with increasing $t_{\mathrm{Sn}}$, the injected spin current from the top Fe layer would be efficiently converted into a charge current, resulting in minimal spin accumulation within the α-Sn layer and reduced backflow spin current. Consequently, a larger net spin current is absorbed by the α-Sn layer, manifesting as the significant enhancement in the Gilbert damping constant $\alpha$. At $t_{\mathrm{Sn}} > 25$ nm, the spin current decays largely in the bulk α-Sn channel before reaching the TSS at the bottom α-Sn/InSb interface, leading to the downfall of $\alpha$. Using $\lambda_{\mathrm{Sn}} = 20 \pm 5$ nm, the experimentally determined inverse Edelstein length $\lambda_{IEE}$ is then estimated to be -3.1±1.0 nm. This value of $\lambda_{IEE}$ is comparable to the largest value reported at room temperature in all materials, which is the case of half-Heusler compound Au/Co/MnPtSb [37], and is achieved without using any spacer layer between Fe and α-Sn. This result establishes TDS α-Sn as one of the most promising platforms for spin-charge interconversion.

It is noteworthy that this estimated value of $\lambda_{IEE}$ is largely affected by the value of $\lambda_{Sn}$, and thus should be taken with caution. For a precise understanding of spin transport and spin–charge interconversion in α-Sn, experimental benchmarking of $\lambda_{Sn}$ using other approaches remains an important future issue. Nevertheless, the extremely high spin-to-charge conversion efficiency and insights into the interplay between TSS evolution and spin-charge interconversion in topological materials advance our understanding of their potential in spintronic applications.

*Acknowledgments*— This work was partly supported by Grants-in-Aid for Scientific Research (19K21961, 20H05650, 22K14591, 23K17324, 24H00018, and 25H00840), CREST program (JPMJCR1777) and NEXUS (JPMJNX25D5) of Japan Science and Technology Agency, Spintronics Research Network of Japan (Spin-RNJ), and "Advanced Research Infrastructure for Materials and Nanotechnology in Japan (ARIM)" of the Ministry of Education, Culture, Sports, Science and Technology (MEXT), Grant Number JPMXP24UT0121.

## References

[1] R. Ramaswamy, J. M. Lee, K. Cai, and H. Yang, Recent advances in spin-orbit torques: Moving towards device applications, Appl. Phys. Rev. **5**, 031107 (2018).

[2] V. D. Nguyen, S. Rao, K. Wostyn, and S. Couet, Recent progress in spin-orbit torque magnetic random-access memory, Npj Spintron. **2**, 48 (2024).

[3] Z. Guo, J. Yin, Y. Bai, D. Zhu, K. Shi, G. Wang, K. Cao, and W. Zhao, Spintronics for Energy- Efficient Computing: An Overview and Outlook, Proc. IEEE **109**, 1398 (2021).

[4] S. Manipatruni, D. E. Nikonov, C.-C. Lin, T. A. Gosavi, H. Liu, B. Prasad, Y.-L. Huang, E. Bonturim, R. Ramesh, and I. A. Young, Scalable energy-efficient magnetoelectric spin–orbit logic, Nature **565**, 35 (2019).

[5] J.-C. Rojas-Sánchez et al., Spin to Charge Conversion at Room Temperature by Spin Pumping into a New Type of Topological Insulator: α-Sn Films, Phys. Rev. Lett. **116**, 096602 (2016).

[6] F. Binda, C. O. Avci, S. F. Alvarado, P. Noël, C.-H. Lambert, and P. Gambardella, Spin-orbit torques and magnetotransport properties of α-Sn and β-Sn heterostructures, Phys. Rev. B **103**, 224428 (2021).

[7] S. H. Su, C.-W. Chong, J.-C. Lee, Y.-C. Chen, V. V. Marchenkov, and J.-C. A. Huang, Effect of Cu Intercalation Layer on the Enhancement of Spin-to-Charge Conversion in Py/Cu/Bi2Se3, Nanomaterials **12**, 3687 (2022).

[8] Y. Wang et al., Room temperature magnetization switching in topological insulator-ferromagnet heterostructures by spin-orbit torques, Nat. Commun. **8**, 1364 (2017).

[9] N. H. D. Khang, Y. Ueda, and P. N. Hai, A conductive topological insulator with large spin Hall effect for ultralow power spin–orbit torque switching, Nat. Mater. **17**, 808 (2018).

[10] S. V. Eremeev, V. N. Men'shov, V. V. Tugushev, P. M. Echenique, and E. V. Chulkov, Magnetic proximity effect at the three-dimensional topological insulator/magnetic insulator interface, Phys. Rev. B **88**, 144430 (2013).

[11] C. Lee, F. Katmis, P. Jarillo-Herrero, J. S. Moodera, and N. Gedik, Direct measurement of proximity-induced magnetism at the interface between a topological insulator and a ferromagnet, Nat. Commun. **7**, 12014 (2016).

[12] P. Noel, C. Thomas, Y. Fu, L. Vila, B. Haas, P.-H. Jouneau, S. Gambarelli, T. Meunier, P. Ballet, and J. P. Attané, Highly Efficient Spin-to-Charge Current Conversion in Strained HgTe Surface States Protected by a HgCdTe Layer, Phys. Rev. Lett. **120**, 167201 (2018).

[13] J. Polaczyński et al., 3D topological semimetal phases of strained α -Sn on insulating substrate, Mater. Today **75**, 135 (2024).

[14] M. R. Scholz et al., Topological surface state of α−Sn on InSb(001) as studied by photoemission, Phys. Rev. B **97**, 075101 (2018).

[15] I. Madarevic et al., Structural and electronic properties of the pure and stable elemental 3D topological Dirac semimetal α-Sn, APL Mater. **8**, 031114 (2020).

[16] Y. Ohtsubo, P. L. Fèvre, F. Bertran, and A. Taleb-Ibrahimi, Dirac Cone with Helical Spin Polarization in Ultrathin α-Sn(001) Films, Phys. Rev. Lett. **111**, 216401 (2013).

[17] S. M. Young, S. Zaheer, J. C. Y. Teo, C. L. Kane, E. J. Mele, and A. M. Rappe, Dirac Semimetal in Three Dimensions, Phys. Rev. Lett. **108**, 140405 (2012).

[18] B.-J. Yang and N. Nagaosa, Classification of stable three-dimensional Dirac semimetals with nontrivial topology, Nat. Commun. **5**, 4898 (2014).

[19] S. Kobayashi and M. Sato, Topological Superconductivity in Dirac Semimetals, Phys. Rev. Lett. **115**, 187001 (2015).

[20] G. M. Stephen, A. T. Hanbicki, T. Schumann, J. T. Robinson, M. Goyal, S. Stemmer, and A. L. Friedman, Room-Temperature Spin Transport in Cd3As2, ACS Nano **15**, 5459 (2021).

[21] W. Yanez et al., Spin and Charge Interconversion in Dirac-Semimetal Thin Films, Phys. Rev. Appl. **16**, 054031 (2021).

[22] J. Ding et al., Switching of a Magnet by Spin-Orbit Torque from a Topological Dirac Semimetal, Adv. Mater. **33**, 2005909 (2021).

[23] C.-Z. Xu et al., Elemental Topological Dirac Semimetal: α-Sn on InSb(111), Phys. Rev. Lett. **118**, 146402 (2017).

[24] K. H. M. Chen et al., Thickness-dependent topological phase transition and Rashba-like preformed topological surface states of α-Sn(001) thin films on InSb(001), Phys. Rev. B (2022).

[25] Q. Barbedienne et al., Angular-resolved photoemission electron spectroscopy and transport studies of the elemental topological insulator α-Sn, Phys. Rev. B **98**, 195445 (2018).

[26] See Supplemental Material for the details of sample preparation, measurements, x-ray diffraction, x-ray reflectivity, transport measurement, and SQUID characterization; Details of ferromagnetic resonance and spin pumping experiment; Details of DFT calculation; Derivation of the theoretical model, which includes Refs. [5,7,12,21,25,29,31,33,35–59]

[27] A. Leis, M. Schleenvoigt, V. Cherepanov, F. Lüpke, P. Schüffelgen, G. Mussler, D. Grützmacher, B. Voigtländer, and F. S. Tautz, Lifting the Spin-Momentum Locking in Ultra-Thin Topological Insulator Films, Adv. Quantum Technol. **4**, (2021).

[28] J. Linder, T. Yokoyama, and A. Sudbø, Anomalous finite size effects on surface states in the topological insulator Bi2Se3, Phys. Rev. B **80**, 205401 (2009).

[29] C. S. Ho, Y. Wang, Z. B. Siu, S. G. Tan, M. B. A. Jalil, and H. Yang, Effect of surface state hybridization on current-induced spin-orbit torque in thin topological insulator films, Sci. Rep. **7**, 792 (2017).

[30] H.-Z. Lu, W.-Y. Shan, W. Yao, Q. Niu, and S.-Q. Shen, Massive Dirac fermions and spin physics in an ultrathin film of topological insulator, Phys. Rev. B **81**, 115407 (2010).

[31] L. D. Anh, K. Takase, T. Chiba, Y. Kota, K. Takiguchi, and M. Tanaka, Elemental Topological Dirac Semimetal α-Sn with High Quantum Mobility, Adv. Mater. **33**, 2104645 (2021).

[32] J. Kim, Y.-T. Chen, S. Karube, S. Takahashi, K. Kondou, G. Tatara, and Y. Otani, Evaluation of bulk-interface contributions to Edelstein magnetoresistance at metal/oxide interfaces, Phys. Rev. B **96**, 140409 (2017).

[33] T. Chiba, S. Takahashi, and G. E. W. Bauer, Magnetic-proximity-induced magnetoresistance on topological insulators, Phys. Rev. B **95**, 094428 (2017).

[34] T. Chiba, A. O. Leon, and T. Komine, Voltage-control of damping constant in magnetic-insulator/topological-insulator bilayers, Appl. Phys. Lett. **118**, 252402 (2021).

[35] P. Deorani, J. Son, K. Banerjee, N. Koirala, M. Brahlek, S. Oh, and H. Yang, Observation of inverse spin Hall effect in bismuth selenide, Phys. Rev. B **90**, 094403 (2014).

[36] V. Sharma, W. Wu, P. Bajracharya, D. Q. To, A. Johnson, A. Janotti, G. W. Bryant, L. Gundlach, M. B. Jungfleisch, and R. C. Budhani, Light and microwave driven spin pumping across FeGaB–BiSb interface, Phys. Rev. Mater. **5**, 124410 (2021).

[37] E. Longo, A. Markou, C. Felser, M. Belli, A. Serafini, P. Targa, D. Codegoni, M. Fanciulli, and R. Mantovan, Giant Spin-Charge Conversion in Ultrathin Films of the MnPtSb Half-Heusler Compound, Adv. Funct. Mater. **34**, 2407968 (2024).

[38] G. Kresse and J. Hafner, Ab initio molecular dynamics for liquid metals, Phys. Rev. B **47**, 558 (1993).

[39] W. M. Haynes, D. R. Lide, and T. J. Bruno, *CRC Handbook of Chemistry and Physics*, 97th ed. (2016).

[40] J. Cheng et al., Coherent Picture on the Pure Spin Transport between Ag/Bi and Ferromagnets, Phys. Rev. Lett. **129**, 097203 (2022).

[41] G. Kresse and J. Furthmüller, Efficient iterative schemes for ab initio total-energy calculations using a plane-wave basis set, Phys. Rev. B **54**, 11169 (1996).

[42] S. P. Dash, S. Sharma, R. S. Patel, M. P. de Jong, and R. Jansen, Electrical creation of spin polarization in silicon at room temperature, Nature **462**, 491 (2009).

[43] S. L. Dudarev, G. A. Botton, S. Y. Savrasov, C. J. Humphreys, and A. P. Sutton, Electron-energy-loss spectra and the structural stability of nickel oxide: An LSDA+U study, Phys. Rev. B **57**, 1505 (1998).

[44] Z. Shi, X. Wang, C. Xu, P. Wang, Y. Liu, and T.-C. Chiang, First-principles study of the topological surface states of α-Sn(111), Phys. Lett. A **384**, 126782 (2020).

[45] Perdew, Burke, and Ernzerhof, Generalized Gradient Approximation Made Simple., Phys. Rev. Lett. **77**, 3865 (1996).

[46] A. Glavic and M. Björck, GenX 3: the latest generation of an established tool, J. Appl. Crystallogr. **55**, 1063 (2022).

[47] A. N. Engel, C. P. Dempsey, H. S. Inbar, J. T. Dong, S. Nishihaya, Y. Chang, A. V. Fedorov, M. Hashimoto, D. Lu, and C. J. Palmstrøm, Growth and characterization of α-Sn thin films on In- and Sb-rich reconstructions of InSb(001), Phys. Rev. Mater. **8**, 044202 (2024).

[48] Y. Zhao, Y. Hu, L. Liu, Y. Zhu, and H. Guo, Helical States of Topological Insulator Bi2Se3, Nano Lett. **11**, 2088 (2011).

[49] J. Ding, C. Liu, Y. Zhang, V. Kalappattil, R. Yu, U. Erugu, J. Tang, H. Ding, H. Chen, and M. Wu, Large Damping Enhancement in Dirac-Semimetal–Ferromagnetic-Metal Layered Structures Caused by Topological Surface States, Adv. Funct. Mater. **31**, (2021).

[50] T. Chiba, G. E. W. Bauer, and S. Takahashi, Magnetization damping in noncollinear spin valves with antiferromagnetic interlayer couplings, Phys. Rev. B **92**, 054407 (2015).

[51] K. Ando and E. Saitoh, Observation of the inverse spin Hall effect in silicon, Nat. Commun. **3**, 629 (2012).

[52] S. E. Barnes, J. Ieda, and S. Maekawa, Rashba Spin-Orbit Anisotropy and the Electric Field Control of Magnetism, Sci. Rep. **4**, 4105 (2014).

[53] X. Tao et al., Self-consistent determination of spin Hall angle and spin diffusion length in Pt and Pd: The role of the interface spin loss, Sci. Adv. **4**, eaat1670 (2018).

[54] J.-C. Rojas-Sánchez, N. Reyren, P. Laczkowski, W. Savero, J.-P. Attané, C. Deranlot, M. Jamet, J.-M. George, L. Vila, and H. Jaffrès, Spin Pumping and Inverse Spin Hall Effect in Platinum: The Essential Role of Spin-Memory Loss at Metallic Interfaces, Phys. Rev. Lett. **112**, 106602 (2014).

[55] Y. Tserkovnyak, A. Brataas, and G. E. W. Bauer, Spin pumping and magnetization dynamics in metallic multilayers, Phys. Rev. B **66**, 224403 (2002).

[56] M. Caminale, A. Ghosh, S. Auffret, U. Ebels, K. Ollefs, F. Wilhelm, A. Rogalev, and W. E. Bailey, Spin pumping damping and magnetic proximity effect in Pd and Pt spin-sink layers, Phys. Rev. B **94**, 014414 (2016).

[57] S. Küfner, J. Furthmüller, L. Matthes, M. Fitzner, and F. Bechstedt, Structural and electronic properties of α-tin nanocrystals from first principles, Phys. Rev. B **87**, 235307 (2013).

[58] H. Wang, J. Kally, J. S. Lee, T. Liu, H. Chang, D. R. Hickey, K. A. Mkhoyan, M. Wu, A. Richardella, and N. Samarth, Surface-State-Dominated Spin-Charge Current Conversion in Topological-Insulator–Ferromagnetic-Insulator Heterostructures, Phys. Rev. Lett. **117**, 076601 (2016).

[59] S. Fukuoka, L. D. Anh, M. Ishida, T. Hotta, T. Chiba, Y. Kota, and M. Tanaka, Topological surface states induced by magnetic proximity effect in narrow-gap semiconductor alpha-Sn, arXiv (2025).

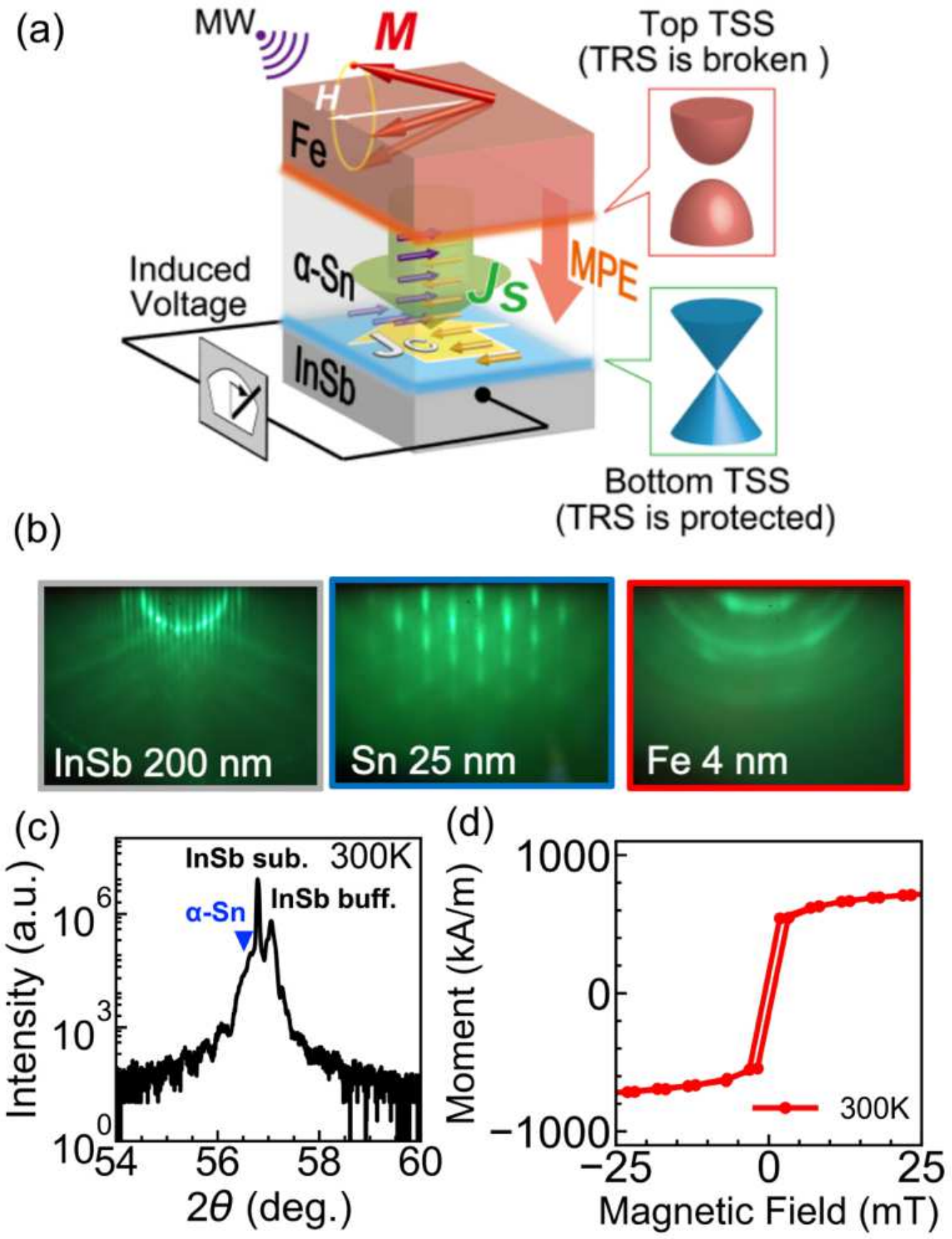

**Fig. 1.** (a) Illustration of the concept and method of this study. In bilayers of topological Dirac semimetal (TDS) and a ferromagnet (Fe), the topological surface states (TSSs) on both sides of TDS are influenced by the magnetic proximity effect (MPE) from the adjacent magnetization. This leads to the breaking of time reversal symmetry (TRS) and suppression of spin-momentum locking. By tuning the TDS layer thickness, the TSS on the opposite interface away from the Fe layer (bottom TSS) remains intact and works as a high-efficiency spin-charge converter. (b) *In situ* RHEED patterns along the [110] direction, (c) X-ray diffraction (XRD) spectrum and (d) Magnetization hysteresis (*M*–*H*) curve at room temperature for the Fe (4 nm) / α-Sn (25 nm) / InSb heterostructure sample (S25).

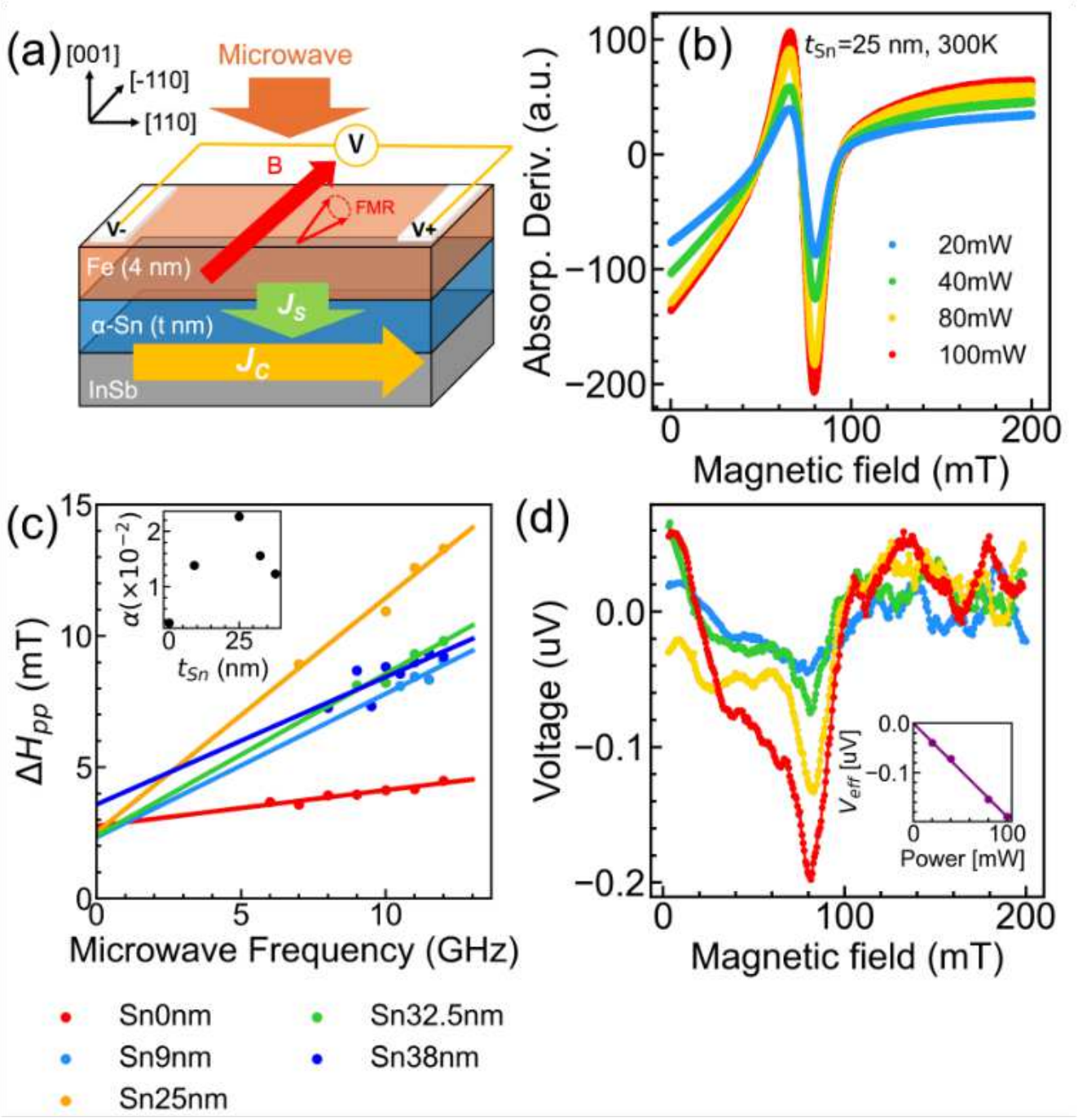

**Fig. 2.** (a) Schematic sample structure and experimental setup for RF frequency-dependent ferromagnetic resonance (FMR) measurements. The heterostructure was attached to a coplanar waveguide, where a RF magnetic field was generated using a vector network analyzer (VNA). (b) RF power-dependent FMR spectra of the Fe (4 nm) / α-Sn (25 nm) / InSb heterostructure sample. (c) RF frequency dependence of the peak-to-peak linewidth $\Delta H_{\mathrm{pp}}$ of the FMR spectrum. The slope of the linear fit yields the Gilbert damping constant α of the system. Inset shows the α-Sn layer thickness dependence of α. (d) RF power dependence of the induced voltage signal in the same sample. The induced voltage is the difference between the voltage signals when the magnetic field is applied along the $[\bar{1}10]$ and $[1\bar{1}0]$ direction (See Fig. S7 in Supplemental Material [26]). Inset shows the RF power dependence of the effective voltage peak $V_{\mathrm{eff}}$, estimated using the procedure described in Supplemental Material.

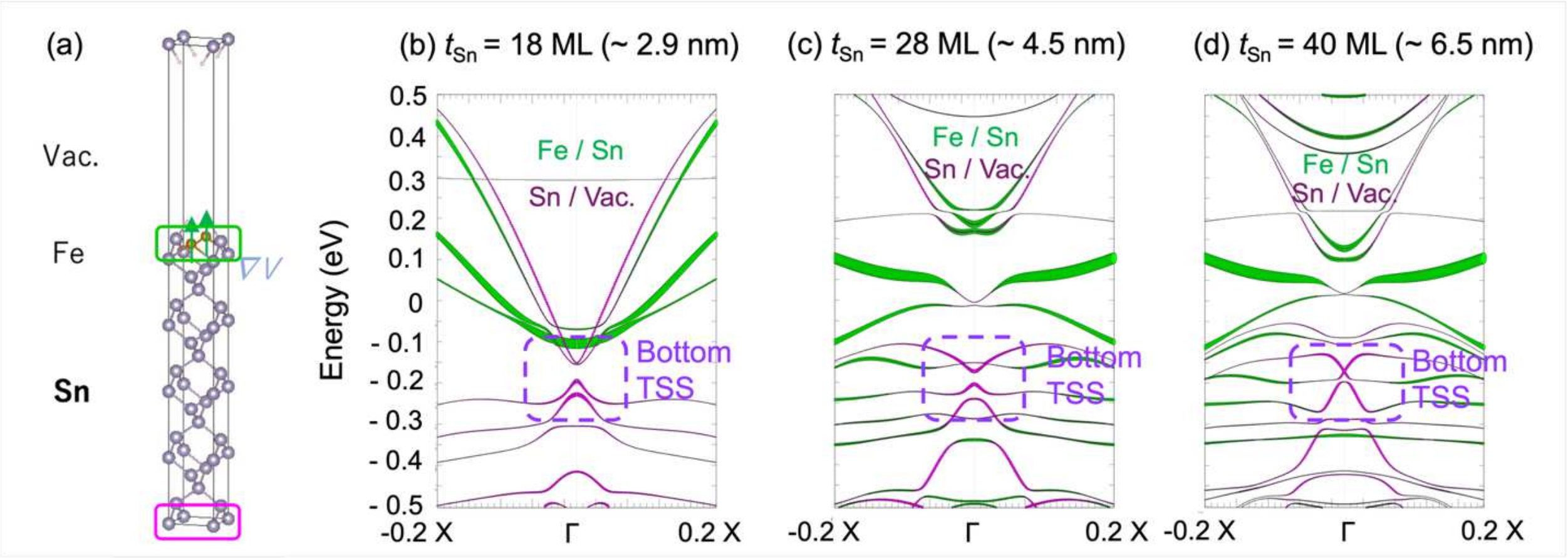

Fig. 3. (a) Crystal structure model for density functional theory (DFT) calculation of the band structure of Fe/α-Sn ($t_{Sn}$ = 18, 28, 40 monolayers (ML), corresponding to ~2.9, 4.5, and 6.5 nm)/vacuum (InSb) heterostructures. (b,c,d) Projection of the band components contributed from the Sn atoms at the interfaces with the top Fe layer (green) and the bottom vacuum (purple). The surface states at the vacuum side (corresponding to the bottom α-Sn/InSb interface in our experiment) are concentrated in the topological gap (surrounded by the purple dashed curves). The TSS exhibits a small exchange gap due to the MPE, which decreases with increasing $t_{Sn}$.

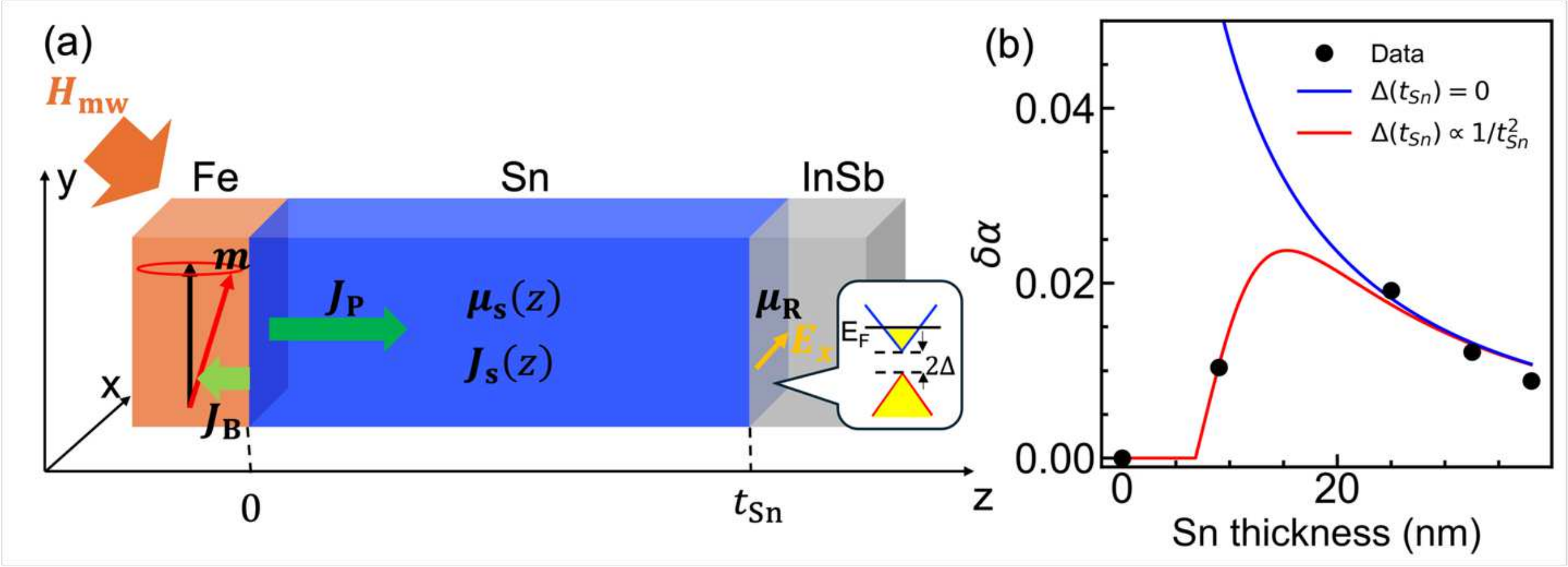

Fig. 4. (a) Schematic view of the Fe/α-Sn/InSb heterostructure and model used in this work. (b) α-Sn layer thickness dependence of the damping enhancement fitted by the proposed model. The blue line shows the fitting result when $\Delta = 0$, and red line shows the fitting result when $\Delta \propto 1/t_{\mathrm{Sn}}{}^2$.

SUPPLEMENTAL MATERIAL

# Giant Spin-to-Charge Conversion by Tailoring Magnetically Proximitized Topological Dirac Semimetal

*Masayuki Ishida[1], Soichiro Fukuoka[1], Takahiro Chiba[2,3]†††, Yohei Kota[4], Masaaki Tanaka[1,5,6]†† and Le Duc Anh[1,5]†*

[1]Dept. of Electrical Engineering and Information Systems, The University of Tokyo, Tokyo, Japan
[2]Department of Information Science and Technology, Graduate School of Science and Engineering, Yamagata University, Japan
[3]Department of Applied Physics, Graduate School of Engineering, Tohoku University, Japan
[4]National Institute of Technology, Fukushima College, Japan
[5]Center for Spintronics Research Network, The University of Tokyo, Tokyo, Japan
[6]Institute for Nano Quantum Information Electronics, The University of Tokyo, Tokyo, Japan

## I. Sample Growth

After an InSb (001) substrate was introduced into the III-V growth chamber of our MBE system, it was annealed at ~400°C to remove the topmost oxide layer. A 100–200 nm-thick InSb buffer layer was then grown at a rate of 500 nm/h on the substrate to ensure an atomically flat surface. The InSb layers of all samples were terminated by an In-rich surface, which was confirmed by the c(8x2) RHEED pattern (see Fig. 1(b) in the main text and Fig. S1). The sample was then transferred into a metal growth chamber without breaking the vacuum, and it was cooled down to below –10°C. α-Sn and Fe layers were grown at a rate of 1 nm/min and 0.25 nm/min, respectively. The RHEED pattern of the α-Sn layer showed streaky 2-fold reconstruction in the [110] direction for all samples, ensuring successful epitaxial growth. The samples were immediately capped by a 3–5 nm Al layer to prevent oxidation.

## II. Sample Characterization

Fig. S1 shows the *in-situ* RHEED pattern of Fe (4 nm) / Sn ($t_{Sn}$ nm) / InSb buffer / InSb substrate heterostructure ($t_{Sn}$ = 0, 9, 25, 32.5, and 38 nm) taken during the MBE growth. InSb layers of all samples were terminated by In-rich, c(8x2) reconstruction following our previous work [S1]. Sn layers in all samples show streaky (2x2) reconstruction, ensuring the growth of diamond-type (α-phase) Sn [S2]. The dominance of the α-phase in the Sn layer is also supported by the *ex-situ* XRD spectrum shown in Fig. S2.

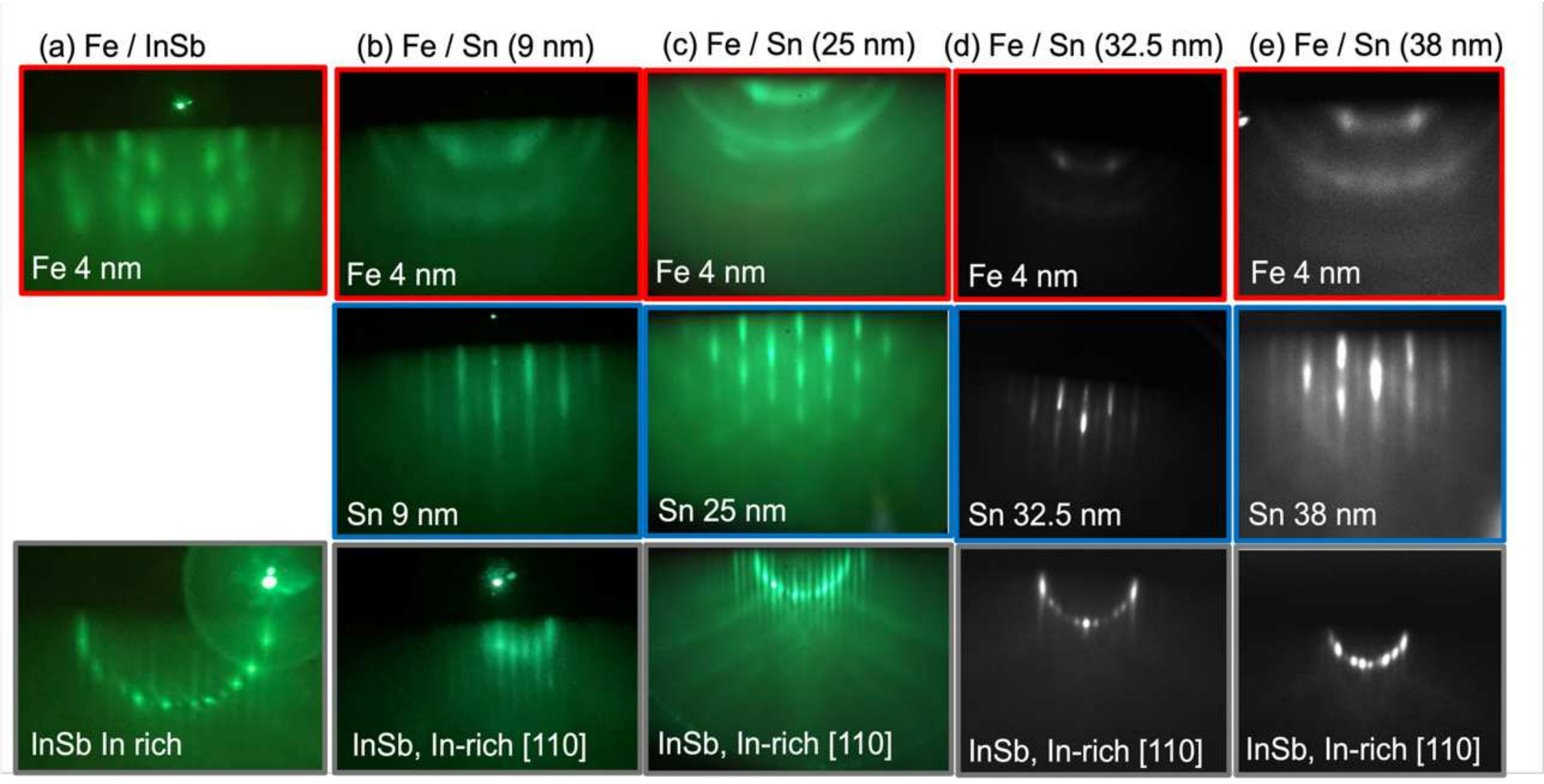

**Fig. S1** (a,b,c,d,e) Reflection high energy electron diffraction (RHEED) patterns along [110] direction of InSb susbtrate during the MBE growth of Fe (4 nm) /α-Sn ($t_{Sn}$ nm) / InSb buffer / InSb ($t_{Sn}$ = 0, 9,25, 32.5, and 38 nm), respectively. All samples with $\alpha$-Sn showed streaky (2x2) reconstruction, ensuring successful epitaxial growth of the α-Sn layer.

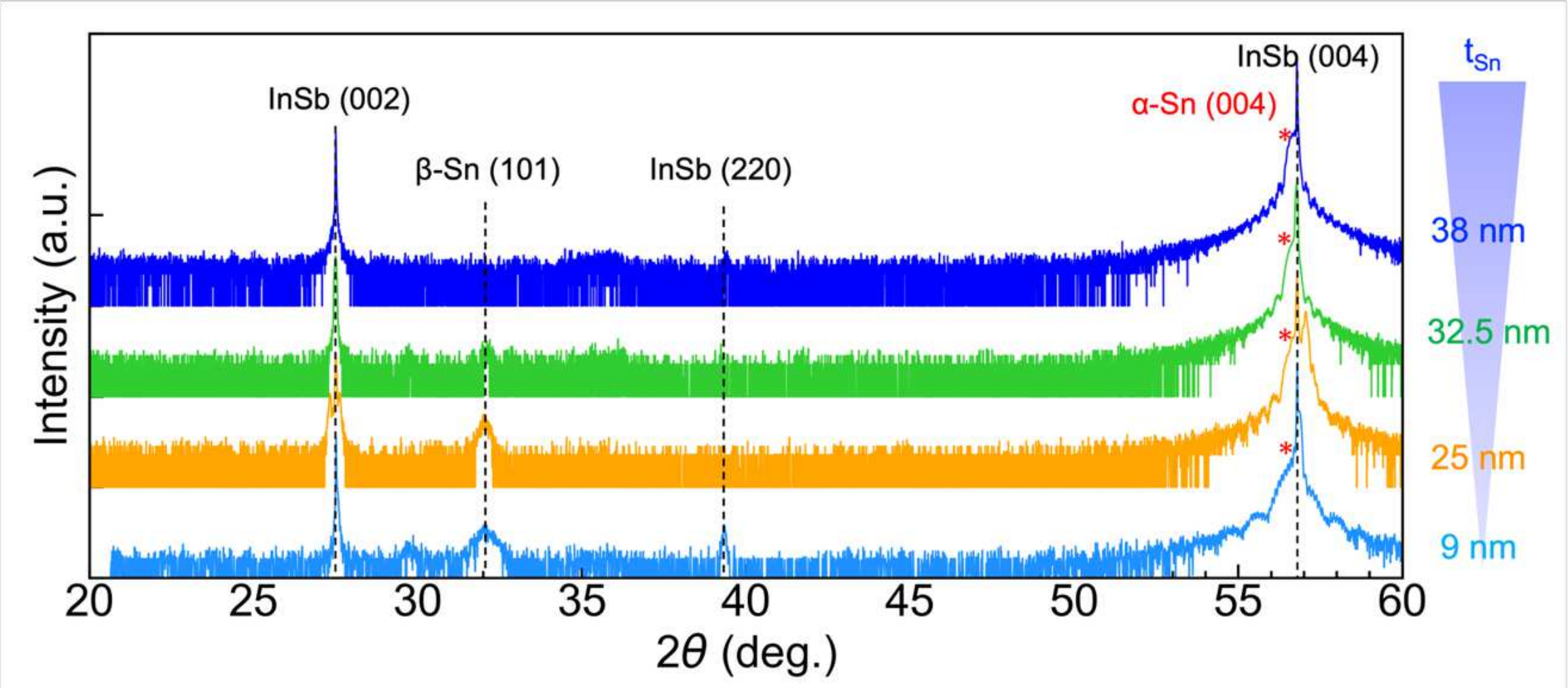

**Fig. S2** X-ray diffraction (XRD) spectra of Fe (4 nm) / Sn ($t_{Sn}$ nm) / InSb buffer / InSb heterostructures ($t_{Sn}$ = 9, 25, 32.5, and 38 nm) observed using Rigaku Samartlab (with 9kW anode X-ray source and HyPix-3000 detector). Clear α-Sn (004) peaks were observed in all samples, accompanied by Laue fringes which were used to estimate the thickness of the α-Sn layer. The presence of β-Sn is negligible in all samples as seen from the small β-Sn peaks compared with the one corresponding to α-Sn.

## III. XRR measurement

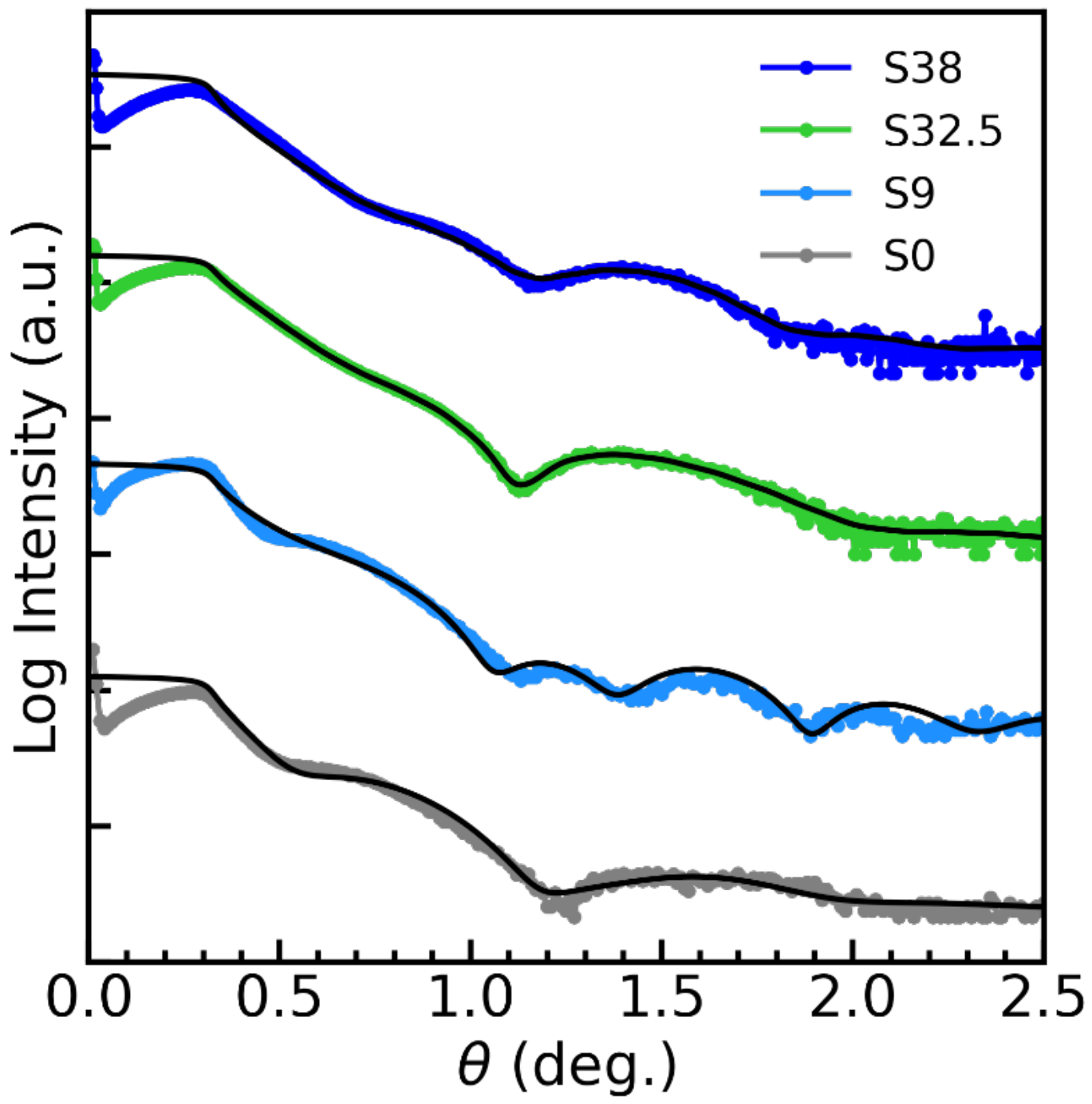

**Fig. S3** X-ray reflectivity measurement on S0, S9, S32.5, and S38. Black solid line shows the fitting result, which was carried out using GenX [S3]. All samples showed Kiessig fringes corresponding to Fe and Al layers. The thickness of Fe and Al capping layers estimated from the fitting is summarized in Table S-I.

**TABLE S-I.** Al and Fe layer thickness of S0, S9, S32.5, and S38 extracted from the XRR measurement.

| Sample ID | Fe layer thickness (nm) | Al layer thickness (nm) |
|---|---|---|
| S0 | 4.0 | 5.5 |
| S9 | 4.3 | 5.3 |
| S32.5 | 4.2 | 3.4 |
| S38 | 4.24 | 3.387 |

## IV. Transport Measurement

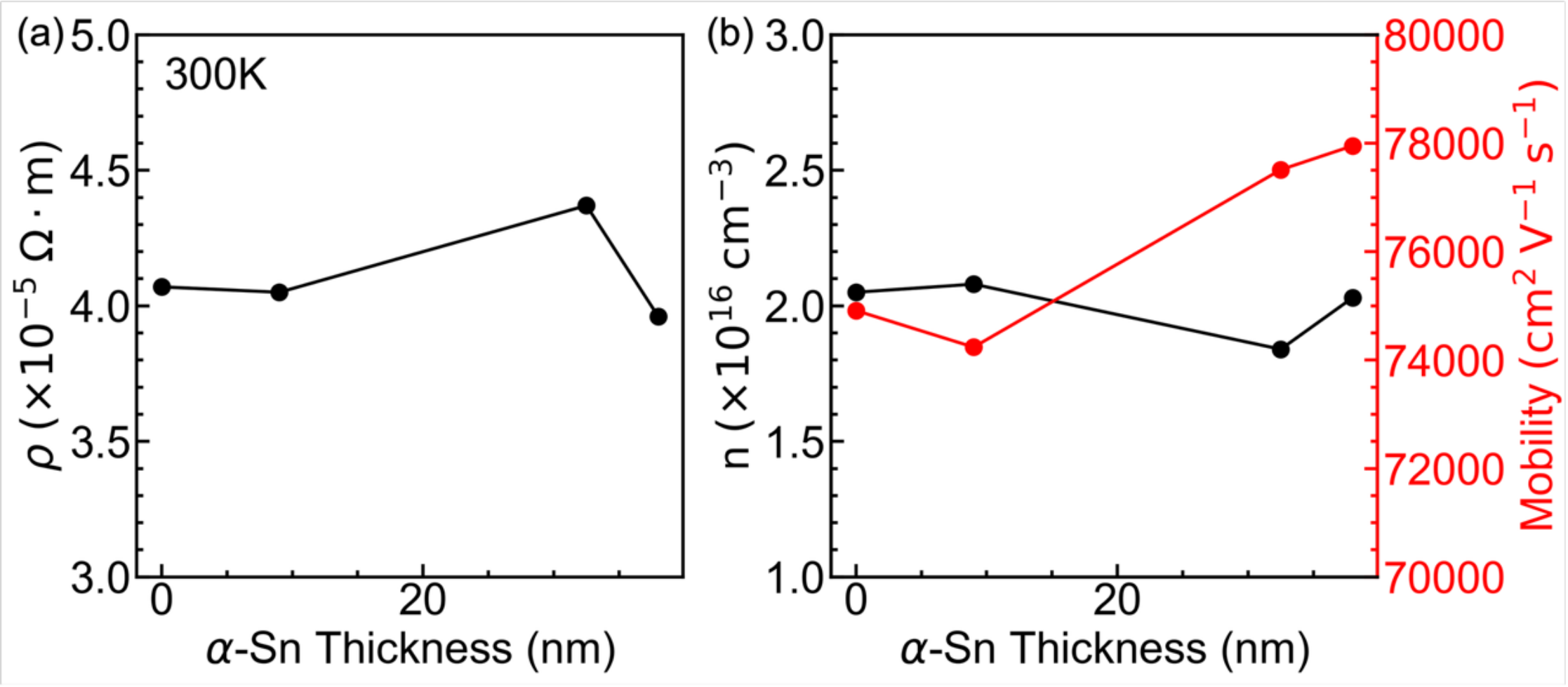

**Fig. S4** (a) Resistivity ($\rho$), (b) carrier density ($n$), and mobility of Fe (4 nm) / α-Sn (0, 9, 32.5, 38 nm) / InSb heterostructure estimated from the Van der Pauw method. All the samples showed n-type conduction, indicating dominant parallel conduction in the InSb substrate.

In order to convert the observed voltage into charge current, the resistivity, carrier density, and mobility of samples were measured using the Van der Pauw method at 300K. The results are shown in **Fig. S4**, where the resistivity showed little α-Sn thickness dependence due to massive current shunting in the InSb buffer and substrate layers, which are both conductive. The existence of parallel conduction in InSb is also supported by the little α-Sn thickness dependence of carrier density and mobility, in addition to the fact that dominant carriers observed here were electrons (n-type conduction), whereas p-type conduction is expected for α-Sn since the Fermi level of α-Sn lies within its heavy hole band [S1].

Although characterizing transport properties of only the α-Sn layer in our present samples is challenging, we do not expect significant changes in the bulk transport properties of α-Sn for thicknesses above 9 nm. This is supported by our previous work [S1], where transport in α-Sn (3.4 – 9.2 nm)/InSb heterostructures was investigated using Shubnikov–de Haas (SdH) oscillation analysis. The SdH frequency of the heavy hole (HH)-band oscillation, which reflects the bulk Fermi surface size, remains unchanged once the α-Sn thickness exceeds 40 ML (~6.4 nm). This indicates that the bulk transport properties of α-Sn remain essentially unchanged for $t_{Sn} > 6.4$ nm, where α-Sn is in the topological Dirac semimetal phase. Since all the samples used in this work (except the reference sample without α-Sn) contain α-Sn layers thicker than 9 nm, it is therefore reasonable to assume that their bulk transport properties do not vary significantly.

## V. SQUID Measurement

**Fig. S5** shows the *M-H* curves of Fe (4 nm) / Sn ($t_{Sn}$ nm) / InSb buffer / InSb heterostructures ($t_{Sn}$ = 0, 9, 25, 32.5, and 38 nm) measured using a SQUID magnetometer. MPMS3 by Quantum Design was used to carry out the measurement.

All samples showed clear hysteresis in *M-H* curves both at room temperature (300K) and lower temperature (2K or 4K). No shift in hysteresis or significant change in coercive force was observed in any of the samples even at lower temperature, indicating negligible effect of exchange bias between Fe and Sn layers.

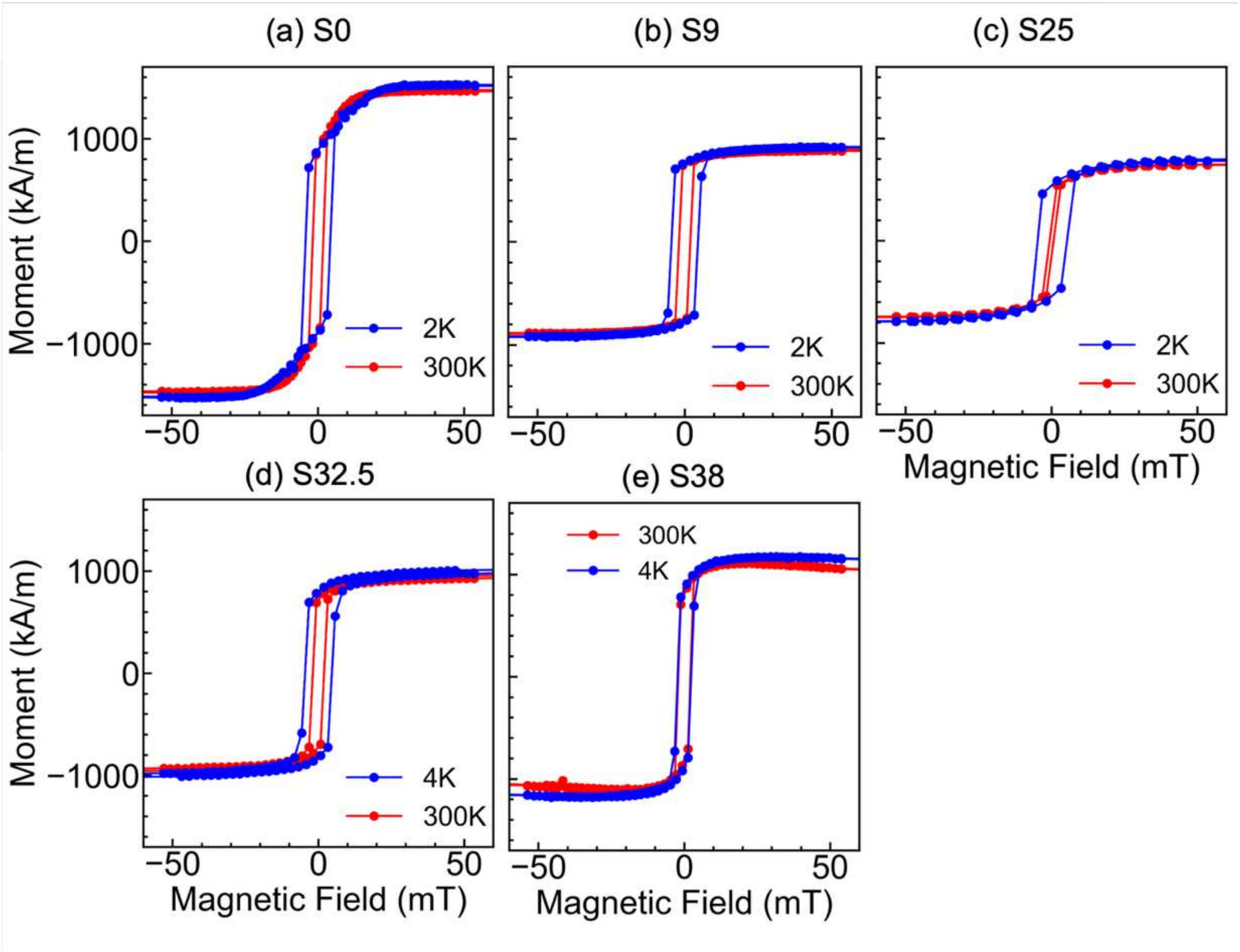

**Fig. S5** (a,b,c,d,e) *M-H* curves of Fe (4 nm) / Sn ($t_{Sn}$ nm) / InSb buffer / InSb heterostructures ($t_{Sn}$ = 0, 9, 25, 32.5, and 38 nm, labeled here as S0, S9, S25, S32.5, and S38) obtained from SQUID magnetometry.

## VI. Ferromagnetic Resonance Measurements Results

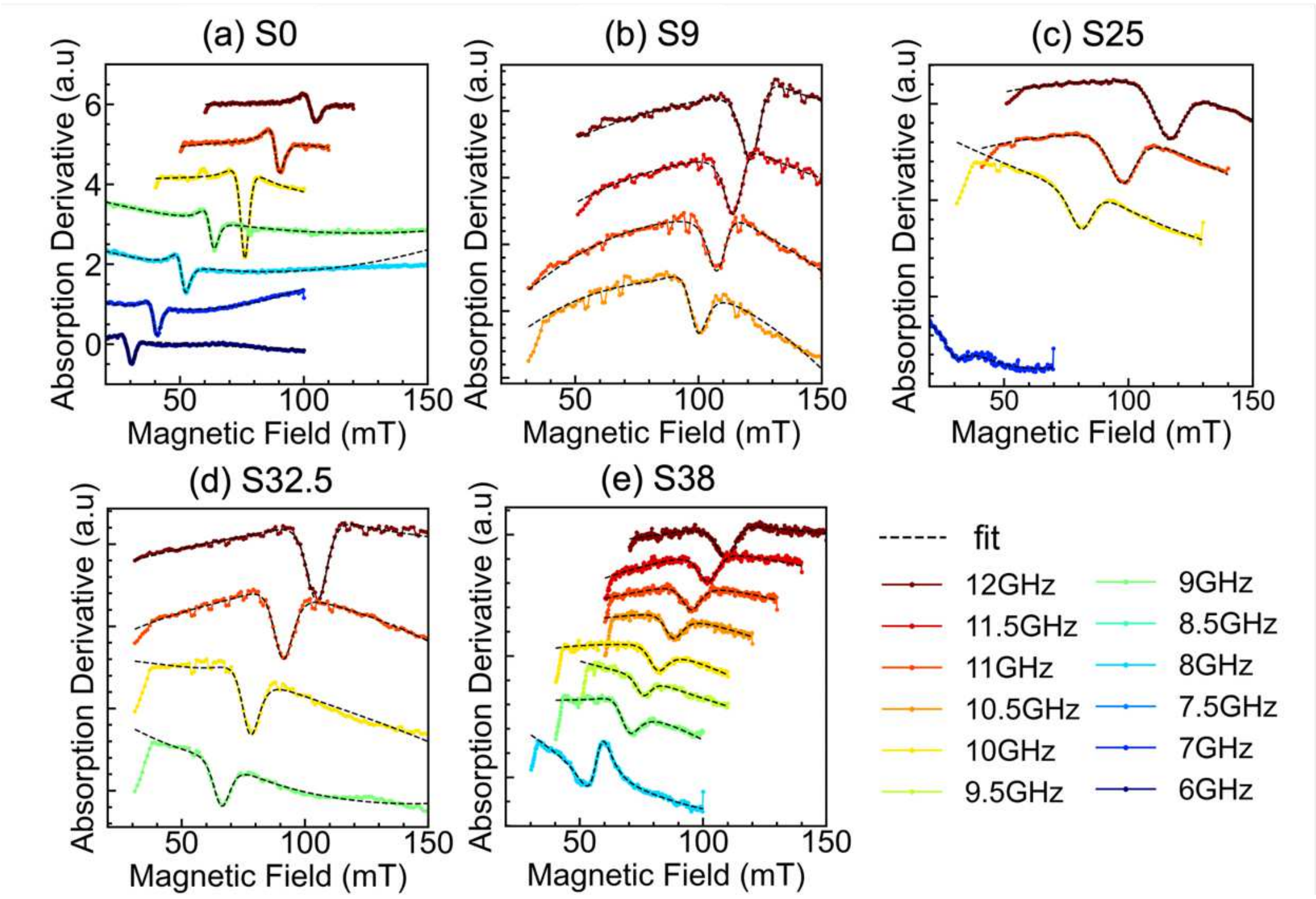

**Fig. S6** (a,b,c,d,e) The raw FMR spectra for S0, S9, S25, S32.5, and S38, respectively. A constant offset, proportional to the RF frequency under which each data was obtained, was added to each spectrum. Black dotted lines are the fitting curves using equation S1. All FMR spectra were measured under a RF power of 10 mW.

The ferromagnetic resonance (FMR) spectra of S0, S9, S25, S32.5, and S38 used to estimate the Gilbert damping constant are shown in **Fig. S6**. The absorption derivative was estimated using the magnetic-field dependence of the $S_{21}$ parameter of the coplanar waveguide attached to the sample using a Keysight VNA. Data of S0 show spectra with the shape similar to Lorentzian derivative, yet for S9, S25, S32.5, and S38, the spectra show the shapes similar to that of anti-Lorentzian derivative. However, the FMR field extracted from the data follows the Kittel equation with reasonable saturation magnetization and anisotropy field, indicating that the observed signals originate from FMR. The abnormal sspectra shapes are most likely due to the measurement setup and possibly induced by the lack of phase-matching circuit in the setup.

The FMR spectra were fitted with a linear combination of Lorentzian and anti-Lorentzian functions and a quadratic background component, expressed in the following equation

$$\frac{dS_{21}}{dH} = -\frac{2LX}{\Delta H(X^2+1)^2} - \frac{A(1-X^2)}{\Delta H(X^2+1)^2}$$
$$+C_2H^2 + C_1H + C_0 \qquad (S1)$$

where $\frac{dS_{21}}{dH}$ is the magnetic field derivative of RF absorption intensity, $X$ is a parameter defined as $X = \frac{H-H_r}{\Delta H}$, $\Delta H = \frac{\sqrt{3}}{2}\Delta H_{\mathrm{pp}}$, $H$ is the applied magnetic field, $H_r$ is the resonance field, and $\Delta H_{\mathrm{pp}}$ is the peak-to-peak linewidth of the resonance spectrum. $L$ and $A$ are the magnitudes of the Lorentzian and anti-Lorentzian terms, respectively. $C_0$, $C_1$, and $C_2$ are the constants of the polynomial fitting part.

For the estimation of the Gilbert damping constant, the information of the full-width half-maximum of the RF absorption is required. Although the data shown in Fig. S6(b-e) show the spectra similar to the shape of anti-Lorentzian derivative (expressed as the second term of the right side of Eq. (S1)) due to phase offset between the RF magnetic field driving FMR and the magnetization of the sample itself, this phase offset should not affect the overall full-width half-maximum of the RF absorption. Since the width of the apparent dip in the anti-Lorentzian derivative also corresponds to the half-width half-maximum of the RF absorption and the peak-to-peak linewidth of Lorentzian-derivative-like FMR spectra, we concluded that these data are viable in estimating the Gilbert damping constant.

The frequency dependence of the peak-to-peak linewidth of the resonant field $\Delta H_{\mathrm{pp}}$ is analyzed to rule out the inhomogeneous broadening of the FMR spectrum. The relation between $\Delta H_{\mathrm{pp}}$ (in the unit of mT) and the Gilbert damping constant $\alpha$ can be expressed by Eq. (S2) [S4]

$$\Delta H_{\mathrm{pp}} = \frac{4\pi\alpha}{\sqrt{3}\gamma} f + \Delta H_{\mathrm{pp0}} \qquad (S2)$$

where $f$ is the frequency of the applied microwave in the unit of Hz, $\gamma = 1.84 \times 10^{11}\ \mathrm{T^{-1}s^{-1}}$ is the gyromagnetic ratio when g-factor of Fe ($g = 2.09$) is assumed, $\Delta H_{\mathrm{pp0}}$ is the inhomogeneity-induced broadening in the unit of mT. From the experimental values of $\Delta H_{\mathrm{pp}}$, the Gilbert damping constant $\alpha$ was estimated.

The enhancement of the Gilbert damping constant $\alpha$ is positively correlated to the increase of the mixing conductance and injected spin current from Fe into α-Sn. The real part of the mixing conductance $g_r (= 2e^2G_r/h)$ in the unit of $\mathrm{m^{-2}}$ can be expressed as a function of the enhancement of the Gilbert damping constant as in Eq. (S3) [S5,S6],

$$g_r = \frac{4\pi M_s d_M}{g\mu_B}\left(\alpha_{\mathrm{Fe|Sn|InSb}} - \alpha_{\mathrm{Fe|InSb}}\right) \tag{S3}$$

where $M_s$ is the saturation magnetization of the sample in the unit of kA/m, $d_M$ is the thickness of the Fe layer in the unit of meter, $g = 2.09$ is Landé's g factor of Fe, $\mu_B = 9.27 \times 10^{-24}$ J/T is the Bohr magneton, and $\alpha_{\mathrm{Fe|Sn|InSb}}$ and $\alpha_{\mathrm{Fe|InSb}}$ are the Gilbert damping constants of samples with and without α-Sn layer, respectively. The spin mixing conductance is then related to the injected spin current $J_s$ at the Fe/α-Sn interface ($z = 0$) as shown in Eq. (S4).

$$J_s(0) = \frac{g_r \gamma^2 h_{\mathrm{mw}}^2 \left[\mu_0 M_s \gamma + \sqrt{(\mu_0 M_s \gamma)^2 + 4\omega^2}\right]}{8\pi \alpha_{\mathrm{Fe|Sn|InSb}}^2 \left[(\mu_0 M_s \gamma)^2 + 4\omega^2\right]} \tag{S4}$$

Here, $\mu_0 = 4\pi \times 10^{-7}\ \mathrm{T \cdot m/A}$ is the vacuum permeability, and $h_{mw}$ and $\omega$ are the strength (in the unit of mT) and angular frequency (in the unit of rad/s) of the RF field, respectively. From these equations, the mixing conductance $g_r$ is estimated to be $54 \pm 8$ $\mathrm{nm}^{-2}$ and the injected spin current is estimated to be 3.2 MA $\mathrm{m}^{-2}$ when applied RF power was 100 mW. At the α-Sn/InSb interface, the spin current is reduced as

$$J_s(t_{Sn}) = J_s(0) e^{-\frac{t_{Sn}}{\lambda_{Sn}}} \tag{S5}$$

where $\lambda_{Sn} = 25$ nm is the upper limit of spin diffusion length of Sn estimated in section X and $t_{Sn} = 25$ nm is the thickness of Sn layer of S25.

## VII. Spin Pumping Measurement

We fitted the following equation to the electromotive force voltage $V_{\mathrm{EMF}}$ measured in the spin-pumping experiments,

$$\begin{aligned} V_{EMF} &= V_{\mathrm{sym}} \frac{\Delta H^2}{(H - H_r)^2 + \Delta H^2} \\ &+ V_{\mathrm{asym}} \frac{\Delta \mathrm{H}(H - H_r)}{(H - H_r)^2 + \Delta H^2} \\ &+ D_3 H^3 + D_2 H^2 + D_1 H + D_0 \end{aligned} \tag{S6}$$

where $V_{\mathrm{sym}}$ is the amplitude of the Lorentzian term, $V_{\mathrm{asym}}$ is the amplitude of the anti-Lorentzian term, $H_r$ is the magnetic field of the peak voltage, $\Delta\mathrm{H}$ is the half-width half maximum of the voltage peak, and $D_0$, $D_1$, $D_2$, and $D_3$ are the constants of the polynomial fitting part. $V_{H+}$ and $V_{H-}$ mentioned in the main text take the value of $V_{\mathrm{sym}}$ in the corresponding experiments. The induced charge current was then calculated from $V_{\mathrm{eff}} =$

$(V_{H+} - V_{H-})\ /\ 2$ and the sheet resistance $R_{\mathrm{sheet}}$ as follows:

$$J_c = \frac{V_{\mathrm{eff}}}{R_{\mathrm{sheet}}} \qquad (S7)$$

From the injected spin current $J_s$ and the induced charge current $J_c$, we can estimate the inverse Edelstein length as $J_s/J_c$. This method was also employed by Rojas-Sánchez *et al.* [S6], but we modified the definition of $J_s$: Instead of taking it as the spin current at the Fe/α-Sn interface as defined in ref. [S6], we define $J_s(t_{\mathrm{Sn}})$ as the spin current at the α-Sn/InSb interface. This reflects our scenario, where the dominant source of SCC originates from the TSS at the α-Sn/InSb interface.

## VIII. Spin pumping results of Fe/α-Sn/InSb heterostructures

**Fig. S7** presents the induced voltage signals measured in spin pumping experiments on samples with $t_{\mathrm{Sn}} = 25$ nm. The sample was placed in the cavity of an ESR machine, where it was exposed to microwaves at a frequency of approximately 9.1 GHz, while the magnetic field was swept along the $[\bar{1}10]$ and $[1\bar{1}0]$ directions. A polynomial background was subtracted from the raw data to produce the plotted signals. The quantities $V_{H+}$ and $V_{H-}$ in the main text refer to the peak voltages observed under a magnetic field applied along the $[\bar{1}10]$ and $[1\bar{1}0]$ direction, respectively. Ideally, the spin-to-charge conversion (SCC) should produce voltages of equal magnitude but opposite in sign. However, in the Fe/α-Sn samples, we observed positive values for both $V_{H+}$ and $V_{H-}$. This anomaly is likely due to parasitic effects independent of magnetic field direction, namely Seebeck effect. Notably, a similar behavior was reported in ref. [S6]. The intrinsic voltage induced by spin-to-charge conversion was thus extracted as half the difference between $V_{H+}$ and $V_{H-}$.

**Fig. S8** shows spin pumping result at room temperature of S0, S9, S32.5, S38, and **Fig. S9** shows the RF power dependence of intrinsic voltage induced by spin-to-charge conversion and amplitude of RF absorption derivative. The induced voltages were estimated using the procedure outlined in the section VII. No distinct voltage peaks were observed in S0, suggesting that InSb contribution to the spin-to-charge conversion is negligible. S9 exhibited a voltage peak, yet it remained constant in amplitude as the RF power increased, suggesting a non-spin-related origin. In contrast, S32.5 and S38 showed a voltage peak that increased with applied RF power. However, the behavior of voltage induced in S32.5 is not linear, and the voltage of S38 also started to saturate around 8 mW, despite the RF absorption not saturating. Therefore, this voltage peak in S32.5 and S38 is most likely not due to spin-pumping, and may be due to parasitic effects such as spin rectification and spin-Seebeck effects.

In all these samples, the contribution from spin-to-charge conversion is relatively small and obscured by parasitic effects, resulting in a non-linear dependence of the voltage peak on the RF power. Therefore, to clearly measure the voltage arising from spin-to-charge conversion, it is necessary to optimize the sample structure to enhance the signal relative to parasitic contributions. On the other hand, by measuring the Gilbert damping constant, we can avoid the electrical contribution of parasitic thermoelectric effects. Therefore, Gilbert damping constant can serve as a more reliable indicator of spin injection efficiency.

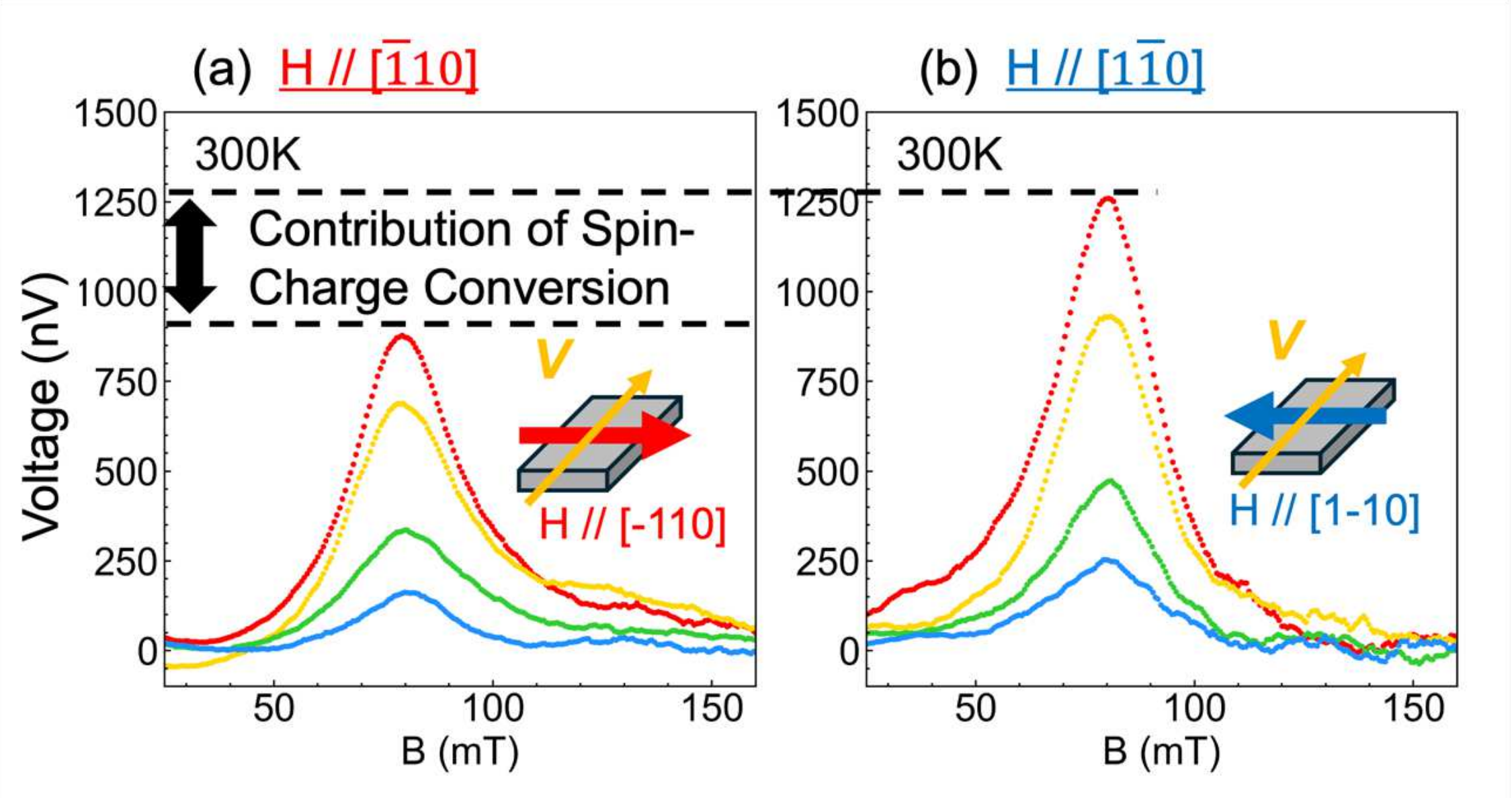

**Fig. S7.** Voltage peak observed in the Fe / α-Sn (25 nm) / InSb heterostructure under sweeping a magnetic field in the (a) $[\bar{1}10]$ and (b) $[1\bar{1}0]$ direction. Polynomial background was subtracted from the raw data to yield the figure. $V_{H+}$ and $V_{H-}$ in the main text correspond to the peak values of the voltage under a magnetic field in the $[\bar{1}10]$ and $[1\bar{1}0]$ direction, respectively.

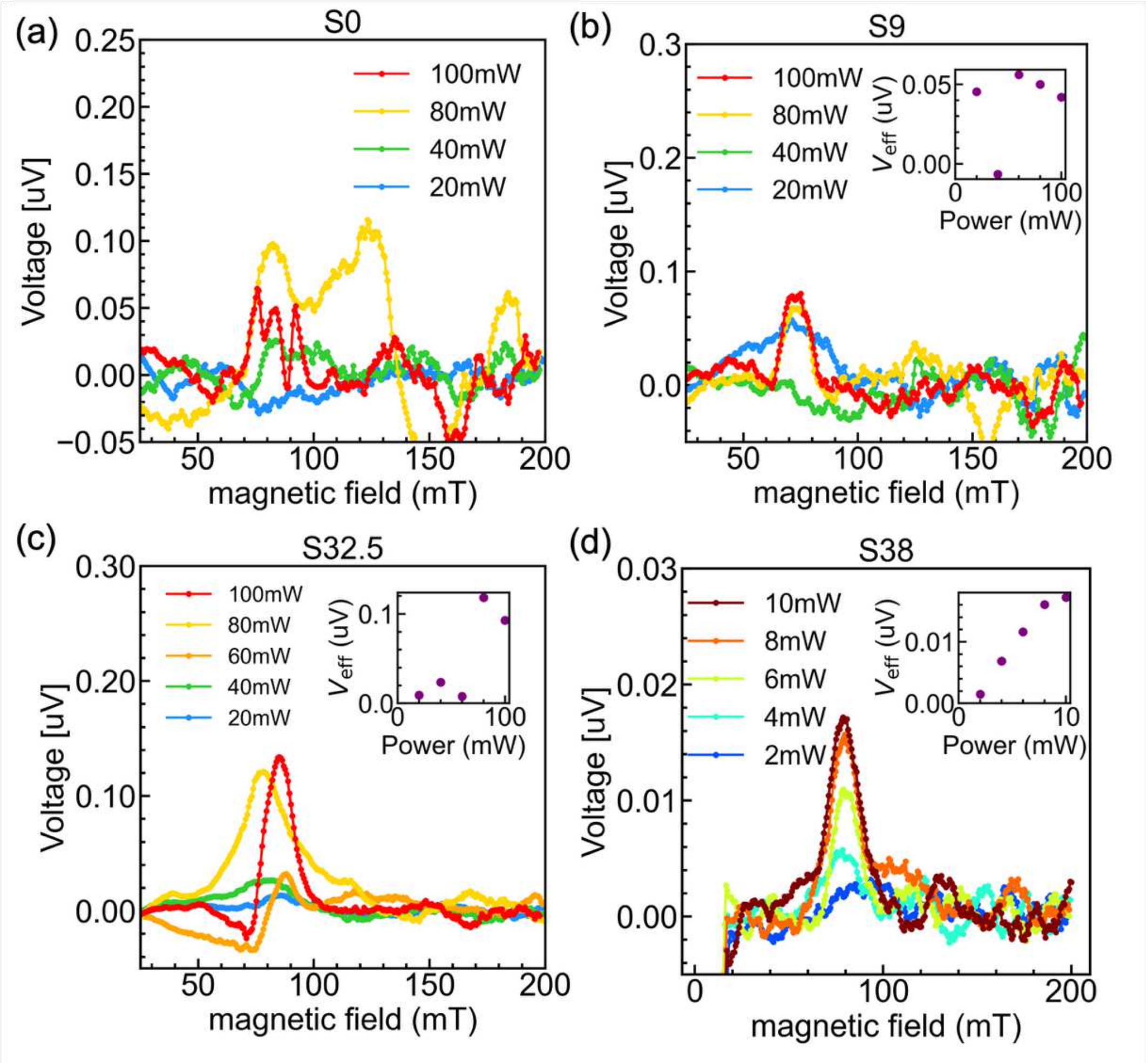

**Fig. S8.** (a,b,c,d) Spin pumping measurement on samples with $t_{Sn}$ = 0, 9, 32.5, 38 nm at room temperature, respectively. Inset in each figure shows the RF power dependence of $V_{eff.}$ The induced voltages were estimated using the same procedure described in section VII. No clear voltage peak was observed for samples with $t_{Sn}$ = 0 nm, while the sample with $t_{Sn}$ = 9 nm showed a voltage peak that did not change in amplitude as the RF power increased. This voltage peak is most likely due to Seebeck effect and the contribution of spin-charge conversion is negligible. The induced voltage peak in the sample with $t_{Sn}$ = 32.5 and 38 nm showed an increasing trend as the applied RF power increased, yet it did not increase linearly with increasing RF power. Measurement of S38 above 20mW showed the saturation of FMR intensity and was therefore not included in this manuscript.

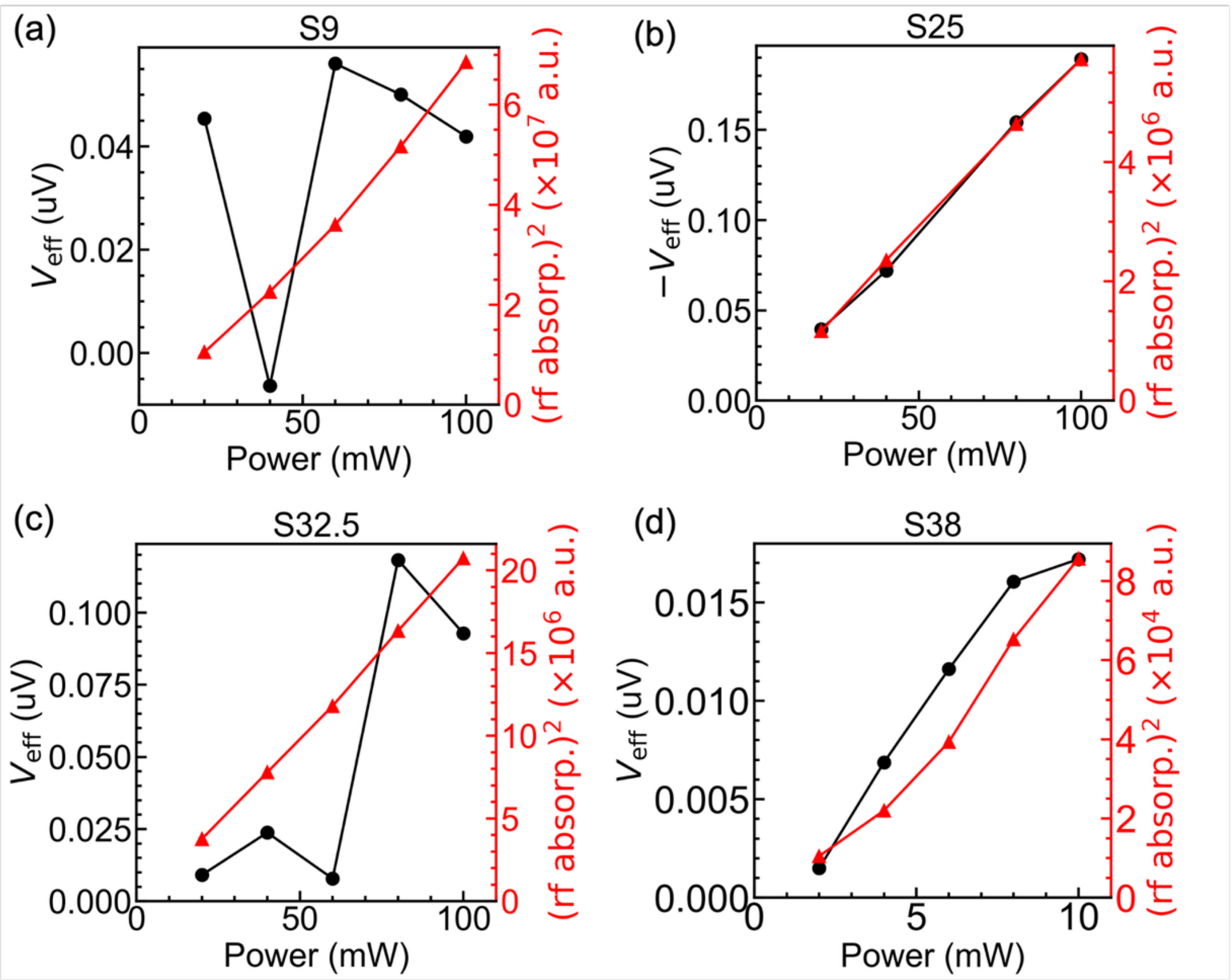

**Fig. S9.** (a,b,c,d) RF power dependence of $V_{\mathrm{eff}}$ and square of the RF absorption derivative intensity in S9, S25, S32.5, and S38, respectively. Among the four samples, only the $V_{\mathrm{eff}}$ of S25 scales linearly with the square of RF absorption amplitude, which is proportional to the effective RF magnetic field strength.

## IX. Details of Density Functional Theory calculation

The first-principles density functional theory calculation was performed by using the projector augmented wave potentials implemented in the Vienna ab initio simulation package (VASP) [S7,S8]. The exchange–correlation energy was treated within the generalized gradient approximation (GGA) by Perdew-Burke-Ernzerhof [S9]. Spin–orbit coupling was taken into consideration in the self-consistent calculations of the electronic band structure. The cutoff energy of the plane wave basis was fixed to 500 eV. The Brillouin zone integration in slab geometry calculations was replaced by a sum over 8 × 8 × 1 Monkhorst-Pack k-point meshes. A lattice constant of α-Sn, 6.4765 Å, was obtained by the total energy minimization of a bulk α-Sn system in the framework of the GGA+U method [S10], where U = –2.5 eV was used for the calculation. This value of U parameter provides a better description of the electronic states in the bulk and slab geometries of α-Sn, such as the topological electronic structure around the band gap and the topological phase transition property [S1,S11–S13].

For the calculations, the model of Fe/α-Sn slab geometry shown in Fig. 3a was constructed by the layered tetragonal unit cell based on the diamond-type crystal structure deposited along the [001] direction. We considered the Sn slab geometry in which a single Fe atom is embedded as a dopant on the topmost Sn layer. Both edges of the layered unit cell were terminated with H atoms to avoid dangling bonds. The surfaces between slabs were separated by a vacuum layer with a thickness of 2 nm. The 18, 28, and 40 monolayers (ML) slab geometry was adopted for the computation of the topological surface state (TSS) of strained α-Sn. A biaxial in-plane compressive strain of –0.76% was considered to take into account the strain effect on the topological electronic states [S1]. Fixing the in-plane lattice constant, structural optimization of the atomic position was performed.

## X. Derivation of the thickness dependence of damping enhancement

In our model illustrated in Fig. S10, the α-Sn layer is treated as a channel with an out-of-plane electrical conductivity $\sigma$ and spin diffusion length $\lambda_{\mathrm{Sn}}$. It should be noted that the spin diffusion length employed here is an effective, spatially averaged parameter along the out-of-plane direction, rather than a macroscopic, spatially uniform value. This approximation was necessary, since the experimental methods employed in the present work do not allow for the spatially resolved determination of the thickness-dependent spin diffusion length. On the other hand, the approximation should be applicable since all the samples investigated here are in the three-dimensional topological Dirac semimetal regime and no major topological phase change is expected in the investigated α-Sn thickness range. The spin accumulation $\boldsymbol{\mu}_\mathrm{s}(z)$ and spin current $\boldsymbol{J}_\mathrm{s}(z)$, where the direction of the vectors represents the spin orientation along the *y* axis and not the flow along the *z* axis, are determined by the following equations.

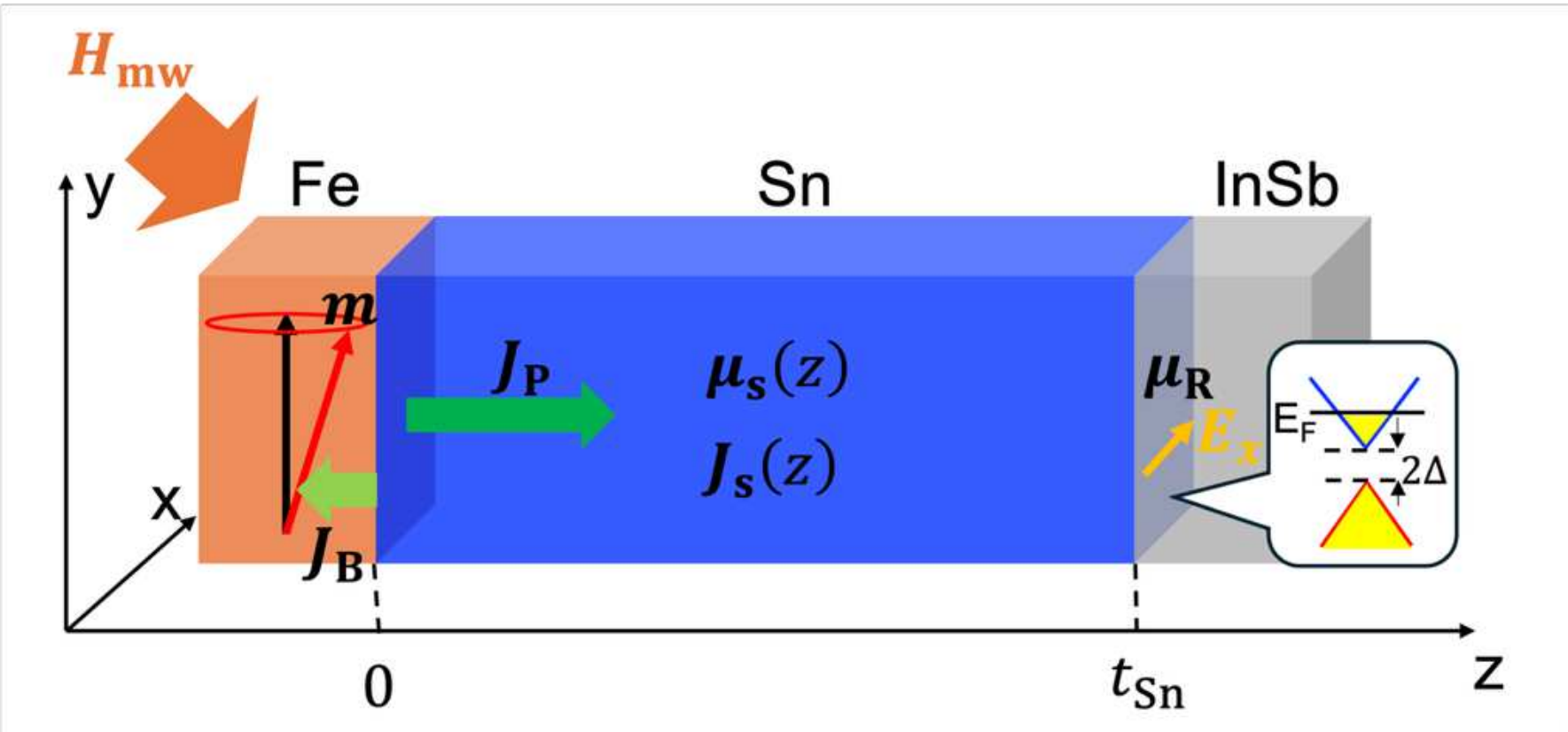

**Fig. S10.** α-Sn bulk is modeled as a spin channel, while the α-Sn/InSb interface is modeled as a gapped TSS with a magnitude of $2\Delta$. The precession of unit magnetization vector $\boldsymbol{m}$ generates a spin current $\boldsymbol{J}_\mathrm{P}$ by spin pumping, while the spin within the α-Sn bulk creates a backflow spin current $\boldsymbol{J}_\mathrm{B}$, leading to the effective injected spin current $\boldsymbol{J}_\mathrm{eff}$.

$$\frac{\partial^2 \boldsymbol{\mu}_\mathrm{s}(z)}{\partial z^2} = \frac{\boldsymbol{\mu}_\mathrm{s}(z)}{\lambda_\mathrm{Sn}^2} \tag{S8}$$

$$\boldsymbol{J}_\mathrm{s}(z) = -\frac{\sigma}{e}\frac{\partial \boldsymbol{\mu}_\mathrm{s}(z)}{\partial z}\ (\parallel y) \tag{S9}$$

On the other hand, in the presence of magnetic proximity effect (MPE), a gapped spin-polarized topological surface state of α-Sn described as $E_\pm = \pm\sqrt{(hv_\mathrm{F}\boldsymbol{k})^2 + \Delta^2}$ is assumed at the bottom α-Sn/InSb interface. Assuming the Fermi level is in the $E_+(k)$ band, the 2D spin accumulation at the α-Sn/InSb interface can be estimated as (see Ref. [S14])

$$|\boldsymbol{\mu}_\mathbf{R}| = A\frac{\hbar}{2}\int \frac{d^2k}{(2\pi)^2} f(k)\langle k,+|s_2|k,+\rangle$$

$$= -\mu_\mathrm{R}^0 \frac{1-\left(\frac{\Delta}{E_\mathrm{F}}\right)^2}{1+3\left(\frac{\Delta}{E_\mathrm{F}}\right)^2} \tag{S10}$$

where $\mu_\mathrm{R}^0 = A\frac{\hbar}{2}\frac{2e\boldsymbol{E}_x}{hv_\mathrm{F}}\frac{E_\mathrm{F}\tau_\mathrm{e}}{\hbar}$, $A$ is the area of α-Sn/InSb interface (taken as $3 \times 0.9\ \mathrm{mm}^2$), $s_2 = \begin{pmatrix} 0 & -i \\ i & 0 \end{pmatrix}$ is the $y$ component of the Pauli matrix for electron spin, $v_F$ is the Fermi velocity (taken as $6 \times 10^5$ m/s [S6]), $E_\mathrm{F}$ is the Fermi energy measured from the Dirac point, $E_x$ is the generated electrical field (taken as 1 mV/m based on the experiment in this work), and $\tau_e$ is the relaxation time of electrons at TSS (taken as 0.15 ps [S15]).

We apply the spin scattering theory [S16,S17] to the top Fe/α-Sn interface in the absence of TSS, where the pumped spin current $\boldsymbol{J}_\mathrm{P}$ and backflow spin current $\boldsymbol{J}_\mathrm{B}$ are expressed by Eqs. (S11) and (S12), respectively.

$$\boldsymbol{J}_\mathrm{P} = \frac{G_r}{e}\hbar \boldsymbol{m} \times \frac{d\boldsymbol{m}}{dt} \quad (S11)$$

$$\boldsymbol{J}_\mathrm{B} = \frac{G_r}{e}\boldsymbol{m} \times (\boldsymbol{\mu}(0) \times \boldsymbol{m}) \quad (S12)$$

Here, $G_r$ is the spin mixing conductance in the units of $\Omega^{-1}\mathrm{m}^{-2}$.

From the boundary conditions at the Fe/α-Sn interface ($z$ = 0) and the α-Sn/InSb interface ($z = t_\mathrm{Sn}$), the spin polarization $\boldsymbol{\mu}_s(z)$ in α-Sn layer can be deduced from Eqs. (S8) and (S9) as

$$\boldsymbol{\mu}_\mathrm{s}(z) = -\boldsymbol{J}_\mathbf{eff}\frac{e\lambda_\mathrm{Sn}}{\sigma}\frac{\sinh\left(\frac{z - t_\mathrm{Sn}}{\lambda_\mathrm{Sn}}\right)}{\cosh\left(\frac{t_\mathrm{Sn}}{\lambda_\mathrm{Sn}}\right)} + \boldsymbol{\mu}_\mathbf{R}\frac{\cosh\left(\frac{z}{\lambda_\mathrm{Sn}}\right)}{\cosh\left(\frac{t_\mathrm{Sn}}{\lambda_\mathrm{Sn}}\right)} \quad (S13)$$

where $\boldsymbol{J}_\mathrm{eff}(\equiv \boldsymbol{J}_\mathrm{p} - \boldsymbol{J}_\mathrm{B})$ is the effective injected spin current, $t_\mathrm{Sn}$ is the thickness of the α-Sn layer, and $\boldsymbol{\mu}_R$ is the spin accumulation at the α-Sn/InSb interface. Hence, the spin accumulation at the Fe/α-Sn interface ($z$ = 0) can be expressed as

$$\boldsymbol{\mu}_\mathrm{s}(0) = \boldsymbol{J}_\mathrm{eff}\frac{e\lambda_\mathrm{Sn}}{\sigma}\tanh\left(\frac{t_\mathrm{Sn}}{\lambda_\mathrm{Sn}}\right) + \frac{\boldsymbol{\mu}_\mathrm{R}}{\cosh\left(\frac{t_\mathrm{Sn}}{\lambda_\mathrm{Sn}}\right)} \quad (S14)$$

Since $\boldsymbol{J}_\mathrm{eff}$ is defined as $\boldsymbol{J}_\mathrm{eff} \equiv \boldsymbol{J}_\mathrm{P} - \boldsymbol{J}_\mathrm{B}$, it can be transformed as follows by taking Eq. (S14) into account.

$$\begin{aligned}\boldsymbol{J}_\mathrm{eff} &= \boldsymbol{J}_\mathrm{P} - \frac{G_r}{e}\boldsymbol{m} \times (\boldsymbol{\mu}_\mathrm{s}(0) \times \boldsymbol{m}) \\ &= \boldsymbol{J}_\mathrm{P} - \boldsymbol{J}_\mathrm{eff}\frac{\lambda_\mathrm{Sn}G_r}{\sigma}\tanh\left(\frac{t_\mathrm{Sn}}{\lambda_\mathrm{Sn}}\right) - \frac{G_r}{e}\frac{1}{\cosh\left(\frac{t_\mathrm{Sn}}{\lambda_\mathrm{Sn}}\right)}\boldsymbol{m} \times (\boldsymbol{\mu}_\mathrm{R} \times \boldsymbol{m}) \quad (S15)\end{aligned}$$

Eq. (S15) can be solved for $\boldsymbol{J}_\mathrm{eff}$, which yields

$$\boldsymbol{J}_{\text{eff}} = \hbar \frac{G_r}{e} \frac{1 + \frac{|\boldsymbol{\mu}_{\text{R}}|}{\hbar} \operatorname{sech}\left(\frac{t_{\text{Sn}}}{\lambda_{\text{Sn}}}\right)}{1 + \frac{\lambda_{\text{Sn}} G_r}{\sigma} \tanh\left(\frac{t_{\text{Sn}}}{\lambda_{\text{Sn}}}\right)} \boldsymbol{m} \times \frac{d\boldsymbol{m}}{dt} \tag{S16}$$

The damping enhancement $\delta\alpha$ reflects the effective injected spin current $\boldsymbol{J}_{\text{eff}}$ in the Landau–Lifshitz–Gilbert equation. The relation can be described as

$$\delta\alpha \boldsymbol{m} \times \frac{d\boldsymbol{m}}{dt} = \frac{\gamma}{M_{\text{s}} d_{\text{M}}} \frac{\hbar}{2e} \boldsymbol{m} \times (\boldsymbol{J}_{\text{eff}} \times \boldsymbol{m}). \tag{S17}$$

By substituting Eq. (S16) into Eq. (S17), we yield

$$\delta\alpha\ (t_{\text{Sn}}) = \frac{G_r \gamma \hbar^2}{2e^2 M_{\text{s}} d_{\text{M}}} \frac{1 + \frac{|\boldsymbol{\mu}_{\text{R}}|}{\hbar} \operatorname{sech}\left(\frac{t_{\text{Sn}}}{\lambda_{\text{Sn}}}\right)}{1 + \frac{\lambda_{\text{Sn}} G_r}{\sigma} \tanh\left(\frac{t_{\text{Sn}}}{\lambda_{\text{Sn}}}\right)} \tag{S18}$$

## XI. Fitting Parameters

The parameter values used to fit the experimental results are shown in **Table S-II**. Here, $M_S$, $g_r$, and $d_m$ are parameters obtained from experiment and $\lambda_{Sn}$, $\sigma$, and $E_{\text{F}}$ are fitting parameters. The thickness of the Fe layer was confirmed to be around 4 nm using XRR measurements (see section III). The saturation magnetization $M_S$ is set around 1000 kA/m, which is slightly above half of the value compared with the bulk value of Fe (~1700 kA/m [S18]). This discrepancy is likely due to the small thickness of the Fe layer, where the surface disorder and possible existence of a magnetic dead layer between the Fe and Al capping layers can have a large effect on magnetization.

As for $g_r$, the real part of the spin mixing conductance, the value obtained in this work ($g_r \sim 54\ \text{nm}^{-2}$) is comparable to the one obtained in a sample with Fe (5 nm) / Ag (2 nm) / α-Sn (5 nm) heterostructure in a previous work [S6]. This value is larger compared with values reported for Py/Pt heterostructures (reported to be around $8-32\ \text{nm}^{-2}$ [S19]) but smaller compared with the value reported for Co/Pt (around $80\ \text{nm}^{-2}$ [S20]).

There are several possible factors that can affect the spin mixing conductance.

First, the different growth methods may have affected the interfacial quality between the ferromagnetic and nonmagnetic layers: The Fe/Sn heterostructures were grown using MBE, which leads to higher interface quality than that of the Py/Pt heterostructures, which were fabricated using dc magnetron sputtering.

Second, the choice of the ferromagnetic layer can also affect the spin mixing conductance. As described in the main manuscript, the effective spin mixing conductance

$g_r$ can be described as:

$$g_r = \frac{4\pi M_s d_M}{g\mu_B}\delta\alpha$$

where $M_s$ is the saturation magnetization of the sample in the unit of A/m, $d_M$ is the thickness of the ferromagnetic layer in the unit of meter, $g$ is Landé's g factor, $\mu_B = 9.27 \times 10^{-24}$ J/T is the Bohr magneton, and $\delta\alpha$ is the Gilbert damping constant enhancement due to the presence of the nonmagnetic layer acting as the spin sink (e.g., a Pt layer). Since $g_r$ includes the parameters such as $M_s$, $d_M$, and Landé's $g$ factor, the spin mixing conductance can depend on the property of the ferromagnetic layer. Especially, saturation magnetization $M_s$ is slightly larger for Fe (~1000 kA/m) compared with Py (~827 kA/m [S19]).

Third, the difference in the magnetic proximity effect (MPE) in the nonmagnetic layer may have influenced the effective spin mixing conductance. It is known that magnetic moments are induced at a few monolayers of Pt near the Py/Pt interface [S19], which enhances the spin mixing conductance due to spin decoherence. On the other hand, long-range MPE is expected at Fe/α-Sn heterostructures (as discussed in the main manuscript) which may enhance the spin mixing conductance.

Since the above-mentioned factors are largely independent of one another, it is also possible that several of them contribute simultaneously.

The Fermi level $E_F$ measured from the Dirac point is estimated to be 25 meV, which is in a similar order to that of $Bi_2Se_3$ [S21], albeit somewhat smaller. The Sn layer's out-of-plane conductance $\sigma$ obtained from the fitting is around $7.3 \times 10^6\ \Omega^{-1} \cdot \mathrm{m}^{-1}$, which is a reasonable value.

The α-Sn-thickness dependence of the energy gap $2\Delta$ (in the unit of meV) used to fit the red line in Fig. 4 of the main text is

$$\Delta(t_{Sn}) = \frac{B}{t_{\mathrm{Sn}}^2} = \frac{1800}{t_{\mathrm{Sn}}^2} \qquad (S19)$$

where $t_{\mathrm{Sn}}$ is the thickness of α-Sn in the unit of nm. This relation was determined from the thickness dependence of the energy gap used in ref. [S14] and the constant $B$ = 1800 meV· $\mathrm{nm}^2$ was determined as a value that satisfies $\Delta$=72 meV when $t_{Sn}$=25 nm, which is about 20 meV larger than FM/Bi2Se3 systems [S14]. We attribute this gap enhancement to MPE, as discussed in the main manuscript. It is true that there is a quantitative discrepancy between the results of the DFT calculation and the analytical model: From the DFT calculations shown in Fig. 3 of the main manuscript, the gap opening $2\Delta$ in the bottom TSS gradually decreases with increasing α-Sn thickness and disappears at the thickness of 6.7 nm, yet in the analytical model, $2\Delta$ is not completely closed even above 9 nm, and the analytical model assumes stronger $\Delta = 1800/t_{\mathrm{Sn}}^2$ dependency. However, this discrepancy may have originated from the fact that only one Fe atom is accounted for per unit slab in the model used for the DFT calculation. In reality, the MPE is expected to be stronger than the DFT calculation results.

We note that first-principles calculation results indicate that a large gap in the TSS of α-Sn opens only under MPE from a neighboring perpendicularly magnetized film [S22]. Although in our experiments the magnetization of Fe is mainly in the film plane, there may also be an out-of-plane interfacial magnetization component induced by the Rashba spin-orbit interaction at the Fe/α-Sn interface as described in ref. [S23].

**TABLE S-II.** Fitting parameter values used to fit our model to the experimental results shown in Fig. 4(b) in the main manuscript. $M_s$ and $d_M$ are the saturation magnetization and thickness of Fe respectively, $\sigma$ and $\lambda_{Sn}$ are conductivity and spin diffusion length of α-Sn respectively, $g_r$ is the real part of the spin mixing conductance in the unit of $m^{-2}$, and $E_F$ is the Fermi level.

| $M_s$ | $d_M$ | $\sigma$ | $\lambda_{Sn}$ | $g_r$ | $E_F$ |
|---|---|---|---|---|---|
| 918 kA/m | 4 nm | $7.3 \times 10^6$ $(\Omega \cdot m)^{-1}$ | $20 \pm 5$ nm | 54 $nm^{-2}$ | 25 meV |

**TABLE S-III.** Inverse Edelstein length and spin diffusion length at room temperature in various material structures. Inverse Edelstein length of material structures marked with * was estimated using the relation $\lambda_{IEE} \sim \theta_{ISHE}\lambda_{SD}$, where $\theta_{ISHE}$ is the inverse spin Hall angle and $\lambda_{SD}$ is the spin diffusion length of the material.

| **Material / structure** | **Inverse Edelstein length $\lambda_{IEE}$ (nm)** | **Spin diffusion length $\lambda_{SD}$ (nm)** |
|---|---|---|
| **Fe/α-Sn (This work)** | **-3.1 ± 1.0** | **20 ± 5** |
| Au/Co/MnPtSb | 3.0 [S24] | - |
| Py/Cu/$Bi_2Se_3$ | 2.7 [S25] | - |
| Ni/Ag/Bi/YIG | -2.5 [S26] | 15.7±2.5 [S26] |
| Fe/Ag/α-Sn | 2.1 [S6] | - |
| Py/HgCdTe/HgTe/CdTe | 2.0 ± 0.5 [S27] | - |
| YIG/$Bi_2Se_3$ | 0.035 ± 0.004 [S28] | 6.2±0.15 [S29] |
| FeGaB/BiSb* | 0.0768 [S30] | 7.68 ± 1.2 [S30] |
| Py/Pt* | 0.24 [S31] | 8.0±0.5 [S31] |
| Py/Pd* | 0.037 [S31] | 7.7 ± 0.5 [S31] |
| Py/p-Si* | 0.031 [S32] | 310 [S33] |

**References**

[S1] L. D. Anh, K. Takase, T. Chiba, Y. Kota, K. Takiguchi, and M. Tanaka, Elemental Topological Dirac Semimetal α-Sn with High Quantum Mobility, Adv. Mater. **33**, 2104645 (2021).

[S2] A. N. Engel, C. P. Dempsey, H. S. Inbar, J. T. Dong, S. Nishihaya, Y. Chang, A. V. Fedorov, M. Hashimoto, D. Lu, and C. J. Palmstrøm, Growth and Characterization of α-Sn Thin Films on In- and Sb-Rich Reconstructions of InSb(001), Phys. Rev. Mater. **8**, 044202 (2024).

[S3] A. Glavic and M. Björck, GenX 3: The Latest Generation of an Established Tool, J. Appl. Crystallogr. **55**, 1063 (2022).

[S4] J. Ding, C. Liu, Y. Zhang, V. Kalappattil, R. Yu, U. Erugu, J. Tang, H. Ding, H. Chen, and M. Wu, Large Damping Enhancement in Dirac-Semimetal–Ferromagnetic-Metal Layered Structures Caused by Topological Surface States, Adv. Funct. Mater. **31**, (2021).

[S5] W. Yanez et al., Spin and Charge Interconversion in Dirac-Semimetal Thin Films, Phys. Rev. Appl. **16**, 054031 (2021).

[S6] J.-C. Rojas-Sánchez et al., Spin to Charge Conversion at Room Temperature by Spin Pumping into a New Type of Topological Insulator: α-Sn Films, Phys. Rev. Lett. **116**, 096602 (2016).

[S7] G. Kresse and J. Hafner, Ab Initio Molecular Dynamics for Liquid Metals, Phys. Rev. B **47**, 558 (1993).

[S8] G. Kresse and J. Furthmüller, Efficient Iterative Schemes for Ab Initio Total-Energy Calculations Using a Plane-Wave Basis Set, Phys. Rev. B **54**, 11169 (1996).

[S9] Perdew, Burke, and Ernzerhof, Generalized Gradient Approximation Made Simple., Phys. Rev. Lett. **77**, 3865 (1996).

[S10] S. L. Dudarev, G. A. Botton, S. Y. Savrasov, C. J. Humphreys, and A. P. Sutton, Electron-Energy-Loss Spectra and the Structural Stability of Nickel Oxide: An LSDA+U Study, Phys. Rev. B **57**, 1505 (1998).

[S11] S. Küfner, J. Furthmüller, L. Matthes, M. Fitzner, and F. Bechstedt, Structural and Electronic Properties of α-Tin Nanocrystals from First Principles, Phys. Rev. B **87**, 235307 (2013).

[S12] Q. Barbedienne et al., Angular-Resolved Photoemission Electron Spectroscopy and Transport Studies of the Elemental Topological Insulator α-Sn, Phys. Rev. B **98**, 195445 (2018).

[S13] Z. Shi, X. Wang, C. Xu, P. Wang, Y. Liu, and T.-C. Chiang, First-Principles Study of the Topological Surface States of α-Sn(111), Phys. Lett. A **384**, 126782 (2020).

[S14] C. S. Ho, Y. Wang, Z. B. Siu, S. G. Tan, M. B. A. Jalil, and H. Yang, Effect of Surface State Hybridization on Current-Induced Spin-Orbit Torque in Thin Topological Insulator Films, Sci. Rep. **7**, 792 (2017).

[S15] T. Chiba, S. Takahashi, and G. E. W. Bauer, Magnetic-Proximity-Induced Magnetoresistance on Topological Insulators, Phys. Rev. B **95**, 094428 (2017).

[S16] Y. Tserkovnyak, A. Brataas, and G. E. W. Bauer, Spin Pumping and Magnetization Dynamics in Metallic Multilayers, Phys. Rev. B **66**, 224403 (2002).

[S17] T. Chiba, G. E. W. Bauer, and S. Takahashi, Magnetization Damping in Noncollinear Spin Valves with Antiferromagnetic Interlayer Couplings, Phys. Rev. B **92**, 054407 (2015).

[S18] W. M. Haynes, D. R. Lide, and T. J. Bruno, *CRC Handbook of Chemistry and Physics*, 97th ed. (2016).

[S19] M. Caminale, A. Ghosh, S. Auffret, U. Ebels, K. Ollefs, F. Wilhelm, A. Rogalev, and W. E. Bailey, Spin Pumping Damping and Magnetic Proximity Effect in Pd and Pt Spin-Sink Layers, Phys. Rev. B **94**, 014414 (2016).

[S20] J.-C. Rojas-Sánchez, N. Reyren, P. Laczkowski, W. Savero, J.-P. Attané, C. Deranlot, M. Jamet, J.-M. George, L. Vila, and H. Jaffrès, Spin Pumping and Inverse Spin Hall Effect in Platinum: The Essential Role of Spin-Memory Loss at Metallic Interfaces, Phys. Rev. Lett. **112**, 106602 (2014).

[S21] Y. Zhao, Y. Hu, L. Liu, Y. Zhu, and H. Guo, Helical States of Topological Insulator Bi2Se3, Nano Lett. **11**, 2088 (2011).

[S22] S. Fukuoka, L. D. Anh, M. Ishida, T. Hotta, T. Chiba, Y. Kota, and M. Tanaka, Topological Surface States Induced by Magnetic Proximity Effect in Narrow-Gap Semiconductor Alpha-Sn, arXiv (2025).

[S23] S. E. Barnes, J. Ieda, and S. Maekawa, Rashba Spin-Orbit Anisotropy and the Electric Field Control of Magnetism, Sci. Rep. **4**, 4105 (2014).

[S24] E. Longo, A. Markou, C. Felser, M. Belli, A. Serafini, P. Targa, D. Codegoni, M. Fanciulli, and R. Mantovan, Giant Spin-Charge Conversion in Ultrathin Films of the MnPtSb Half-Heusler Compound, Adv. Funct. Mater. **34**, 2407968 (2024).

[S25] S. H. Su, C.-W. Chong, J.-C. Lee, Y.-C. Chen, V. V. Marchenkov, and J.-C. A. Huang, Effect of Cu Intercalation Layer on the Enhancement of Spin-to-Charge Conversion in Py/Cu/Bi2Se3, Nanomaterials **12**, 3687 (2022).

[S26] J. Cheng et al., Coherent Picture on the Pure Spin Transport between Ag/Bi and Ferromagnets, Phys. Rev. Lett. **129**, 097203 (2022).

[S27] P. Noel, C. Thomas, Y. Fu, L. Vila, B. Haas, P.-H. Jouneau, S. Gambarelli, T. Meunier, P. Ballet, and J. P. Attané, Highly Efficient Spin-to-Charge Current Conversion in Strained HgTe Surface States Protected by a HgCdTe Layer, Phys. Rev. Lett. **120**, 167201 (2018).

[S28] H. Wang, J. Kally, J. S. Lee, T. Liu, H. Chang, D. R. Hickey, K. A. Mkhoyan, M. Wu, A. Richardella, and N. Samarth, Surface-State-Dominated Spin-Charge Current Conversion in Topological-Insulator–Ferromagnetic-Insulator Heterostructures, Phys. Rev. Lett. **117**, 076601 (2016).

[S29] P. Deorani, J. Son, K. Banerjee, N. Koirala, M. Brahlek, S. Oh, and H. Yang, Observation of Inverse Spin Hall Effect in Bismuth Selenide, Phys. Rev. B **90**, 094403 (2014).

[S30] V. Sharma, W. Wu, P. Bajracharya, D. Q. To, A. Johnson, A. Janotti, G. W. Bryant, L. Gundlach, M. B. Jungfleisch, and R. C. Budhani, Light and Microwave Driven Spin Pumping across FeGaB–BiSb Interface, Phys. Rev. Mater. **5**, 124410 (2021).

[S31] X. Tao et al., Self-Consistent Determination of Spin Hall Angle and Spin Diffusion Length in Pt and Pd: The Role of the Interface Spin Loss, Sci. Adv. **4**, eaat1670 (2018).

[S32] K. Ando and E. Saitoh, Observation of the Inverse Spin Hall Effect in Silicon, Nat. Commun. **3**, 629 (2012).

[S33] S. P. Dash, S. Sharma, R. S. Patel, M. P. de Jong, and R. Jansen, Electrical Creation of Spin Polarization in Silicon at Room Temperature, Nature **462**, 491 (2009).